\documentclass[12pt]{article}
\pdfoutput=1
\usepackage[nosort]{cite}

\usepackage{epsfig}
\usepackage{amsfonts}
\usepackage{amscd}
\usepackage{latexsym}
\usepackage{amsmath,amssymb,mathtools}
\usepackage{verbatim}
\usepackage{setspace}
\usepackage[dvipsnames]{xcolor}
\usepackage{fancyhdr}
\usepackage{cite}
\usepackage{hyperref}
\usepackage{xcolor}
\usepackage{tikz}
\usepackage{tikz-cd}
\usepackage{tensor}
\usepackage{slashed}
\usepackage{multirow}
\usetikzlibrary{calc}
\usetikzlibrary{topaths}
\usetikzlibrary{decorations}
\usetikzlibrary{decorations.pathmorphing}
\usetikzlibrary{arrows,decorations.markings,cd}
\usetikzlibrary{calc,arrows,cd,decorations.markings,snakes}
\tikzset{
->-/.style args={#1rotate#2}{decoration={markings, mark=at position #1 with {\arrow[scale=1.5,rotate = #2 ]{stealth}}}, postaction={decorate}}
}
\usetikzlibrary{shapes.geometric}
\usetikzlibrary{knots}
\usepackage{tikz-3dplot}

\tikzset{snake it/.style={decorate, decoration=snake}}

\usepackage{draft}
\usepackage{hyperref}
\usepackage{graphicx,color,subcaption}
\usepackage{cite}
\usepackage{mciteplus}
\usepackage{skak}
\usepackage{bbm}
\usepackage[english]{babel}
\usepackage{amsthm}
\usepackage{soul}
\usepackage{makecell}
 
\numberwithin{equation}{section}

\def\bZ{\mathbb{Z}}

\def\cD{\mathcal{D}}

\def\({\left(}
\def\){\right)}

\newcommand{\thetab}[2]{%
  \theta
  \begin{bmatrix}
    #1 \\
    #2
  \end{bmatrix}%
}

\begin{document}

\begin{titlepage}

\title{Non-invertible Symmetries in Weyl Fermions, and Applications to Fermion-Boundary Scattering Problem}

\author{Pengcheng Wei$^1$ and Yunqin Zheng$^{2}$}

        \address{${}^{1}$ Department of Physics, Nanjing University, Nanjing, Jiangsu 210093, China\\\bigskip 
        ${}^{2}$ School of Quantum and Kavli Institute for Theoretical Sciences, \\University of Chinese Academy of Sciences, Beijing, 100190, China}

\abstract

We construct a family of non-invertible topological defects in two-dimensional theories
of $n$ Weyl fermions.  The construction relies on the existence of $G$-symmetric
conformal boundary conditions for $n$ Dirac fermions.  Upon unfolding, these boundary
conditions become topological defects $\mathcal D$ of $n$ Weyl fermions that intertwine
the two $G$-representations, and they are generically non-invertible.  For
$G=U(1)^n$, we show that $\mathcal D$ is a duality defect associated with gauging a
finite Abelian group $\Gamma$, and we give an explicit algorithm for determining
$\Gamma$ and its action on the fermions.  We also show that the same finite-Abelian
gauging description applies in certain restricted examples with non-Abelian $G$.
By contrast, for certain non-Abelian symmetry structures, including the $G=SU(2)$
symmetry appearing in the $1$-$5$-$7$-$8$-$9$ problem, we prove that $\mathcal D$
cannot be realized as a duality defect for gauging any finite Abelian group.  Finally,
we explain how the duality-defect perspective gives a streamlined derivation of
fermion scattering from a conformal boundary.

\end{titlepage}

\eject

\setcounter{tocdepth}{3} 

\tableofcontents

\section{Introduction}

\subsection{Non-invertible Symmetries in Conformal Field Theories}

It is now widely appreciated that global symmetries in quantum field theories are most naturally formulated in terms of topological defects \cite{Gaiotto:2014kfa}. When such defects obey non-invertible fusion rules, they give rise to what are now called non-invertible symmetries. These symmetries have proved useful in many contexts, including constraints on renormalization group flows and the organization of states and operators. See, for instance, the lecture notes \cite{Bhardwaj:2023kri,Shao:2023gho,Luo:2023ive,Schafer-Nameki:2023jdn,Iqbal:2025dsh,Kaidi:2026urc}. It is therefore natural to ask how many topological defects, and ideally all of them, can be identified in a given quantum field theory.

The simplest quantum field theories for which \emph{all} topological defects are known are the two-dimensional Virasoro minimal models with $c_L=c_R<1$. In these theories, the topological lines are precisely the Verlinde lines \cite{Verlinde:1988sn}.

The next important class consists of the $c_L=c_R=1$ conformal field theories \cite{Ginsparg:1987eb,Ginsparg:1988ui}. Many topological lines in these theories have been constructed explicitly in \cite{Thorngren:2021yso,Fuchs:2007tx}. Some of them are continuous non-invertible symmetries, and have also been discussed from the perspective of non-local currents \cite{Delmastro:2025ksn}. In fact, at the self-dual point of the circle branch, it has been argued that all topological lines are invertible and form an $SO(4)$ group \cite{Moller:2024xtt}. By gauging non-anomalous, possibly continuous, subgroups of this $SO(4)$ symmetry, one can reach all other $c=1$ CFTs \cite{Thorngren:2021yso,Yu:2025iqf,Brandontalk}. In this way, the topological lines of any $c=1$ CFT can be understood systematically.

For CFTs with $c_L=c_R>1$, determining the full set of topological defects remains an open problem. Nevertheless, a distinguished subset is known: for example, the defects that commute with the extended chiral algebra in WZW models.

Another family of particularly simple conformal field theories is provided by chiral CFTs, with $c_R=0$ and $c_L>0$. Such theories have been classified in several low-central-charge regimes. For bosonic chiral CFTs, classifications are available up to $c_L\leq 24$ \cite{DongMason,Schellekens:1992db,vanEkeren:2017scl,Moller:2019tlx,vanEkeren:2020rnz,Hohn:2020xfe}; for fermionic chiral CFTs, analogous low-central-charge classifications are known via fermionization \cite{BoyleSmith:2023xkd,Rayhaun:2023pgc,Hohn:2023auw}. In this note, we focus on the theories of $n$ free Weyl fermions, which are basic building blocks of fermionic chiral CFTs.\footnote{There are other building blocks with larger central charges. } We aim to identify a large class of non-invertible topological defects in these theories, without attempting a complete classification.

\subsection{Folding Trick and Conformal Boundary Conditions}

Consider $n$ 1-component left-moving Weyl fermions $\psi_i(t,x)$, with the standard action
\begin{equation}\label{eq:weylaction}
    S_{\text{Weyl}}= \int d^2x\, \sum_{i=1}^n i \psi_i^\dagger (\partial_t - \partial_x) \psi_i.
\end{equation}
The theory has an evident $U(n)\rtimes \bZ_2$ symmetry: $U(n)$ rotates the fermions, while $\bZ_2$ is charge conjugation, $\psi_i\to \psi_i^\dagger$.

The goal of this note is to ask whether there are additional symmetries beyond $U(n)\rtimes \bZ_2$, and, if so, how to construct them explicitly.

The key observation comes from the folding trick.\footnote{This trick has appeared in several contexts \cite{Tachikawa:2026cxd,Cordova:2025zkz,WatanabeOhmori, Ueda:2025ecm,Antinucci:2025uvj,Furuta:2025ahl}, see also \cite{Watanabetalk,Arias-Tamargo:2026urw,Antinucci:2026uuh}. } Given a putative topological defect $\mathcal{D}$ in a theory of $n$ Weyl fermions, placed along the locus $x=0$, one can fold the theory along $x=0$ to obtain $n$ \emph{Dirac} fermions at the half-plane $x\geq 0$. More concretely, define 
\begin{equation}
\begin{split}
    &\Psi^L_i(t,x):=\psi_i(t,x),  \quad x>0 \\
&\Psi^R_i(t,x):= \psi_i(t,-x), \quad  x<0
\end{split}
\end{equation}
and the action \eqref{eq:weylaction} becomes 
\begin{equation}\label{eq:Diracaction}
    S_{\text{Dirac}}=\int_{x>0} d^2x\left( \sum_{i=1}^n i \Psi_i^{L\dagger} (\partial_t - \partial_x) \Psi_i^L + \sum_{i=1}^n i \Psi_i^{R\dagger} (\partial_t + \partial_x) \Psi_i^R\right).
\end{equation}

The topological line $\mathcal{D}$ in the original theory becomes a conformal boundary condition of the folded theory. Indeed, the fact that $\mathcal{D}$ is topological means that the stress energy tensor is continuous across $x=0$, i.e. $T|_{x\to 0^-}= T|_{x\to 0^+}$. Upon folding, this becomes $(T^L-T^R)|_{x\to 0^+}=0$, which is precisely the Ishibashi condition for a conformal boundary condition.

\begin{figure}
    \centering
    \begin{tikzpicture}[
    scale=1,
    line join=round,
    line cap=round,
    every node/.style={font=\footnotesize}
]

\begin{scope} 
\fill[gray!20]
        (-1,-1) rectangle (1,1);
\draw[decoration = {markings, mark=at position 0.5 with {\arrow[scale=0.8]{stealth}}}, postaction=decorate] (0,-1) -- (0,1);
\node at(0,-1.25){$\mathcal{D}$};
\node at(0.5,0){$\psi$};
\node at(-0.5,0){$\psi$};
\end{scope}

\node at(2.5,0) {$\to$};

\begin{scope}[xshift=5cm]
\fill[gray!40]
        (0,-1) rectangle (1,1);
\draw[decoration = {markings, mark=at position 0.5 with {\arrow[scale=0.8]{stealth}}}, postaction=decorate] (0,-1) -- (0,1);
\node at(0,-1.25){$\bra{\mathcal{D}}$};
\node at(0.6,0){$\begin{pmatrix}
    \Psi^L\\\Psi^R
\end{pmatrix}$};
\end{scope}

\end{tikzpicture}
    \caption{The folding trick. On the left is the theory of Weyl fermions with a topological defect $\cD$. On the right is the theory of Dirac fermions with the conformal boundary condition $\bra{\cD}$.}
    \label{fig:folding trick}
\end{figure}
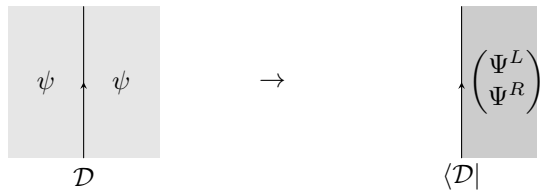

The search for topological defects in a theory of  $n$ Weyl fermions can  therefore be translated to the search for conformal boundary conditions of $n$ Dirac fermions.  There has been both earlier progress\cite{Maldacena:1995pq} and recent progress \cite{Smith:2019jnh,Smith:2020nuf,Smith:2020rru,vanBeest:2023dbu} on the latter problem.  
In general,  these boundary conditions cannot be written as simple linear constraints on $\Psi_i^{L/R}$ at the boundary, and they are opaque in the Lagrangian formulation \eqref{eq:Diracaction}. After unfolding, however, they provide a rich source of  less familiar non-invertible symmetries in the theory of $n$ Weyl fermions.

The conformal boundary conditions discussed in the literature typically preserve a global symmetry, and this symmetry is often chiral: it acts differently on left- and right-moving fermions. Suppose that the Dirac-fermion boundary condition is symmetric under a group $G$ that acts separately on the left- and right-moving fermions:
\begin{align}
    \Psi^L \to U_g\Psi^L U_g^{-1}\quad \Psi^R \to V_g\Psi^R V_g^{-1}\quad \text{for } g\in G,
\end{align}
where the left- and right-moving fermions carry representations $\rho$ and $\widetilde{\rho}$ of $G$, respectively. After unfolding, a $G$-symmetric boundary condition becomes a  topological defect $\cD$ satisfying
\begin{align}\label{symmetric D}
    V_g^\dagger\cD U_g=\cD,\quad \text{or}\quad V_g\cD=\cD U_g,
\end{align}
where $U_g$ acts on $\psi(t,x>0)$ while $V_g$ acts on $\psi(t,x<0)$. Here $\cD$, $U_g$ and $V_g$ are all topological defects in Weyl-fermion theory.

For such $G$-symmetric boundary condition of Dirac fermions to exist, the symmetry $G$ must be anomaly-free \cite{Han:2017hdv,Li:2022drc,Choi:2023xjw,Jensen:2017eof,Kikuchi:2019ytf,Thorngren:2020yht}. In the unfolded Weyl-fermion theory, this means that the anomalies of $U_g$ and $V_g$ match, even though each side may be anomalous by itself.\footnote{Conversely, when $G$ is anomaly free, a $G$-preserving simple conformal boundary condition may not exist. See \cite{Wei:2025zyd} for a recent discussion. }

The main problem addressed in the rest of this paper is the following. Given a conformal boundary condition of $n$ Dirac fermions in 1+1 dimensions that preserves a symmetry $G$, realized by the pair $(V_g, U_g)$, what topological defect $\mathcal{D}$ does it become after unfolding to a theory of $n$ Weyl fermions? Our main results are summarized as follows.
\begin{enumerate}
    \item For abelian $G=U(1)^n$, $\mathcal{D}$ is a duality defect that gauges a finite abelian group $\Gamma$. We give an explicit construction of $\Gamma$ from arbitrary\footnote{Subject to assumptions including anomaly matching on both sides of $\mathcal{D}$.} $U(1)^n$ charge assignments. 
    \item For some but limited examples of non-abelian $G$, such as the theory reduced from $\text{QED}_4$ \cite{vanBeest:2023dbu}, a finite-abelian-group  duality defect still applies.
    \item For a generic non-abelian $G$, such as situations in certain monopoles in the standard model \cite{vanBeest:2023mbs} and the 1-5-7-8-9 problem \cite{vanBeest:2023dbu}, we prove that $\cD$ \emph{cannot} be a duality defect that gauges any finite abelian groups. 
\end{enumerate}

Duality defects are one of the most common examples of non-invertible symmetries \cite{Choi:2021kmx,Kaidi:2021xfk}, and they can be constructed via half space gauging. In recent years, it has become clear that such duality defects are ubiquitous in quantum field theory. Some are related to invertible symmetries by topological manipulations, such as gauging and/or stacking an invertible TQFT, while the others are more intrinsically non-invertible \cite{Kaidi:2022uux,Sun:2023xxv}. Only a few families of the latter type are known, including non-invertible $T$-duality defects in two-dimensional compact bosons, non-invertible $S$-duality defects in four-dimensional Maxwell theories and $\mathcal{N}=4$ super-Yang-Mills theories. The duality defects in 2d Weyl fermions provide another family of this latter kind.

Conversely, the existence of topological defects $\mathcal{D}$ in Weyl fermion theories gives a streamlined,  kinematic  proof of the existence of conformal boundary conditions of $n$ Dirac fermions preserving certain chiral, but non-anomalous, symmetries $G$.

\subsection{Fermion-Boundary Scattering Problem}

One application of the non-invertible symmetries in Weyl fermions is the fermion-boundary scattering problem. This problem was first emphasized by Callan \cite{Callan:1982ac} and Rubakov \cite{Rubakov:1982fp}, who found that, in certain four-dimensional gauge theories, scattering an electron off a magnetic monopole does not  lead to a unitary S-matrix. After an s-wave reduction and after neglecting gauge fluctuation, the problem reduces to a theory of two-dimensional free Dirac fermions on a half-plane, with the monopole worldline becoming an exotic conformal boundary condition. See \cite{vanBeest:2023dbu,Brennan:2023tae,Brennan:2021ewu,vanBeest:2023mbs,Bolognesi:2024kkb} for recent discussions.

Putting these points together, understanding the four-dimensional monopole problem reduces to understanding conformal boundary conditions of two-dimensional Dirac fermions. After unfolding, this further reduces to constructing topological defects in two-dimensional Weyl fermion theories. The last task is precisely the one addressed in this paper.

We emphasize that the interpretation of the monopole worldline as a topological defect in two-dimensional Weyl fermions was worked out in by Kantaro Ohmori and Masataka Watanabe in an unpublished work \cite{WatanabeOhmori} and announced by Wanatabe in an IPMU seminar \cite{Watanabetalk} in 2022. This work studied a particular example with $G= SU(n) \times U(1)$, where the associated topological defect $\mathcal{D}$ is a duality defect that gauges the $\bZ_n$ or $\bZ_{n/2}$ symmetry. Our contribution in the current paper is to systematize this construction for $G=U(1)^n$, and to prove that it cannot be extended to certain more chiral examples with non-abelian $G$.

The outline of the paper is as follows. In Section \ref{sec:intertwining defect}, we formulate the relation between $G$ symmetry $(V_g,U_g)$ and defect $\cD$ in a more general framework, i.e. $\cD$ being a $G$-intertwining defect. We also derive a partition function relation to probe such an intertwining defect.
Section \ref{sec:Defects with Abelian Group Symmetry} is our main result on the abelian group intertwining defect. We explicitly construct them, showing that they are duality defects for finite abelian groups. Using this construction, we come back to the fermion-boundary scattering problem.
In Section \ref{sec:Nonabelian}, we make some remarks on the non-abelian group intertwining defect, providing situations where it \emph{cannot} be realized as such an abelian group duality defect.

\emph{Note Added:} While this paper was being finalized, two related papers \cite{Arias-Tamargo:2026urw,Antinucci:2026uuh} appeared that contain some overlapping results. The paper \cite{Arias-Tamargo:2026urw} classifies duality defects of $n=2$ Weyl fermions, while \cite{Antinucci:2026uuh} discusses the scattering off a boundary using defect anomalies. Our work is complementary; rather than focusing on classification, we broadly explore the duality defects of an arbitrary number of Weyl fermions.

\section{$G$-Intertwining Defect}\label{sec:intertwining defect}

We begin by discussing some general properties of a topological defect $\mathcal{D}$, and implications on the torus partition functions.

\subsection{Properties of Intertwining Defects}

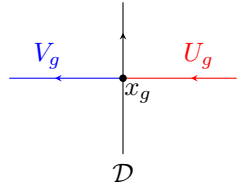
\begin{figure}
    \centering
    \begin{tikzpicture}[
    scale=1,
    line join=round,
    line cap=round,
    every node/.style={font=\footnotesize}
]

\begin{scope}
\draw[decoration = {markings, mark=at position 0.8 with {\arrow[scale=0.8]{stealth}}}, postaction=decorate] (0,-1) -- (0,1);
\draw[red, decoration = {markings, mark=at position 0.4 with {\arrow[scale=0.8]{stealth}}}, postaction=decorate] (1.5,0) -- (0,0);
\draw[blue,decoration = {markings, mark=at position 0.6 with {\arrow[scale=0.8]{stealth}}}, postaction=decorate] (0,0) -- (-1.5,0);
\node[circle, fill=black, inner sep=1pt] at (0,0) {};
\node at(0,-1.25){$\mathcal{D}$};
\node[red] at(1,0.3){$U_g$};
\node[blue] at(-1,0.3){$V_g$};
\node at (0.2,-0.2){\footnotesize{$x_g$}};
\end{scope}
\end{tikzpicture}
    \caption{Topological defect $\cD$ is $G$-intertwining. The defects $U_g$ and $V_g$ can terminate \emph{topologically} on $\cD$ at a common junction.}
    \label{fig: symmetric defect}
\end{figure}

Motivated by the discussion above, we say that the topological defect $\cD$ is $G$-intertwining if 
\begin{align}\label{VD=DU}
    V_g\cD=\cD U_g,
\end{align}
for all $g\in G$. Here, $V_g$ and $U_g$ are topological defects corresponding to a \emph{continuous} group $G$, which need not be abelian. Equivalently, the defects $U_g$ and $V_g$ can terminate \emph{topologically} on $\cD$ at a common junction, as in Figure \ref{fig: symmetric defect}. 

We assume that the fusion $\bar{\cD}\otimes \cD$ yields a \emph{finite} direct sum of simple lines, which we denote by $A$ \cite{Diatlyk:2023fwf}: 
\begin{align}
    \bar{\cD}\otimes \cD=1\oplus\cdots\equiv A.
\end{align}
In fact, we can assign a canonical (co)algebra structure to $A$ via (co)evaluation maps, promoting it to a separable Frobenius algebra. The defect $\cD$ is then the duality defect obtained by gauging the algebra $A$. We remark that this statement applies to a general topological defect, and $\mathcal{D}$ is not necessarily a Tambara-Yamagami defect. 

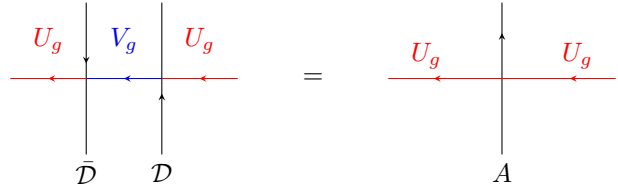
\begin{figure}
    \centering
\begin{tikzpicture}[
    scale=1,
    line join=round,
    line cap=round,
    every node/.style={font=\footnotesize}
]
\begin{scope}
\draw[decoration = {markings, mark=at position 0.4 with {\arrow[scale=0.8]{stealth}}}, postaction=decorate] (-0.5,1) -- (-0.5,-1);
\draw[decoration = {markings, mark=at position 0.4 with {\arrow[scale=0.8]{stealth}}}, postaction=decorate] (0.5,-1) -- (0.5,1);
\draw[red, decoration = {markings, mark=at position 0.5 with {\arrow[scale=0.8]{stealth}}}, postaction=decorate] (1.5,0) -- (0.5,0);
\draw[blue, decoration = {markings, mark=at position 0.5 with {\arrow[scale=0.8]{stealth}}}, postaction=decorate] (0.5,0) -- (-0.5,0);
\draw[red, decoration = {markings, mark=at position 0.5 with {\arrow[scale=0.8]{stealth}}}, postaction=decorate] (-0.5,0) -- (-1.5,0);
\node at(0.5,-1.25){$\mathcal{D}$};
\node at(-0.5,-1.25){$\bar{\mathcal{D}}$};
\node[red] at(1,0.5){$U_g$};
\node[blue] at(0,0.5){$V_g$};
\node[red] at(-1,0.5){$U_g$};
\end{scope}

\begin{scope} [xshift=5cm]
\draw[decoration = {markings, mark=at position 0.8 with {\arrow[scale=0.8]{stealth}}}, postaction=decorate] (0,-1) -- (0,1);
\draw[red, decoration = {markings, mark=at position 0.4 with {\arrow[scale=0.8]{stealth}}}, postaction=decorate] (1.5,0) -- (0,0);
\draw[red,decoration = {markings, mark=at position 0.6 with {\arrow[scale=0.8]{stealth}}}, postaction=decorate] (0,0) -- (-1.5,0);
\node at(0,-1.25){$A$};
\node[red] at(1,0.3){$U_g$};
\node[red] at(-1,0.3){$U_g$};
\end{scope}

\node at(2.5,0) {$=$};

\end{tikzpicture}
    \caption{Transverse intersection of $U_g$ with $\bar{\cD}\otimes\cD=A$.}
    \label{fig: intersect U and A}
\end{figure}

By taking the intersection of $U_g$ and $\bar{\cD}\otimes\cD$, as illustrated in Figure \ref{fig: intersect U and A}, we conclude that $U_g$ commutes with $A$ for any $g\in G$. Because $G$ is a connected continuous group and $A$ consists of a finite sum of simple lines, we further find that\footnote{In particular, $a U_g = U_g b$ can not hold for different $a,b$ in $A$ and for any $g\in G$. This is because we can continuously change $g$ to identity assuming $G$ is connected. When $G$ is not connected, e.g. $G=O(2)$, the following reasoning does not apply. }
\begin{align}\label{aU=Ua}
    aU_g=U_ga, \quad \text{for any } a\subset A.
\end{align}

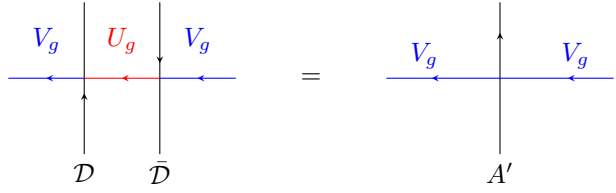
\begin{figure}[t]
    \centering
    \begin{tikzpicture}[
    scale=1,
    line join=round,
    line cap=round,
    every node/.style={font=\footnotesize}
]
\begin{scope}
\draw[decoration = {markings, mark=at position 0.4 with {\arrow[scale=0.8]{stealth}}}, postaction=decorate] (-0.5,-1) -- (-0.5,1);
\draw[decoration = {markings, mark=at position 0.4 with {\arrow[scale=0.8]{stealth}}}, postaction=decorate] (0.5,1) -- (0.5,-1);
\draw[blue, decoration = {markings, mark=at position 0.5 with {\arrow[scale=0.8]{stealth}}}, postaction=decorate] (1.5,0) -- (0.5,0);
\draw[red, decoration = {markings, mark=at position 0.5 with {\arrow[scale=0.8]{stealth}}}, postaction=decorate] (0.5,0) -- (-0.5,0);
\draw[blue, decoration = {markings, mark=at position 0.5 with {\arrow[scale=0.8]{stealth}}}, postaction=decorate] (-0.5,0) -- (-1.5,0);
\node at(0.5,-1.25){$\bar{\mathcal{D}}$};
\node at(-0.5,-1.25){$\mathcal{D}$};
\node[blue] at(1,0.5){$V_g$};
\node[red] at(0,0.5){$U_g$};
\node[blue] at(-1,0.5){$V_g$};
\end{scope}

\begin{scope} [xshift=5cm]
\draw[decoration = {markings, mark=at position 0.8 with {\arrow[scale=0.8]{stealth}}}, postaction=decorate] (0,-1) -- (0,1);
\draw[blue, decoration = {markings, mark=at position 0.4 with {\arrow[scale=0.8]{stealth}}}, postaction=decorate] (1.5,0) -- (0,0);
\draw[blue,decoration = {markings, mark=at position 0.6 with {\arrow[scale=0.8]{stealth}}}, postaction=decorate] (0,0) -- (-1.5,0);
\node at(0,-1.25){$A'$};
\node[blue] at(1,0.3){$V_g$};
\node[blue] at(-1,0.3){$V_g$};
\end{scope}

\node at(2.5,0) {$=$};

\end{tikzpicture}
    \caption{Transverse intersection of $V_g$ with $\cD\otimes\bar{\cD}=A'$.}
    \label{fig: intersect V and Al}
\end{figure}

Similarly, assuming $A'\equiv\cD\otimes\bar{\cD}$ is also a finite sum, we have
\begin{align}\label{aV=Va}
    a'V_g=V_ga', \quad \text{for any } a'\subset A'.
\end{align}
See Figure \ref{fig: intersect V and Al}. Note that since $\bar{\cD}$ is the duality defect obtained by gauging $A'$, the algebra object $A'$ contains the quantum symmetry lines associated with gauging $A$. 

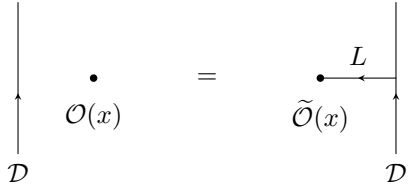
\begin{figure}
    \centering
    \begin{tikzpicture}[
    scale=1,
    line join=round,
    line cap=round,
    every node/.style={font=\footnotesize}
]
\begin{scope}
\draw[decoration = {markings, mark=at position 0.4 with {\arrow[scale=0.8]{stealth}}}, postaction=decorate] (-1,-1) -- (-1,1);
\node[circle, fill=black, inner sep=1pt] at (0,0) {};
\node at(-1,-1.25){$\mathcal{D}$};
\node at(0,-0.5) {$\mathcal{O}(x)$};
\end{scope}

\begin{scope} [xshift=3cm]
\draw[decoration = {markings, mark=at position 0.4 with {\arrow[scale=0.8]{stealth}}}, postaction=decorate] (1,-1) -- (1,1);
\node[circle, fill=black, inner sep=1pt] at (0,0) {};
\draw[decoration = {markings, mark=at position 0.5 with {\arrow[scale=0.8]{stealth}}}, postaction=decorate] (1,0) -- (0,0);
\node at(1,-1.25){$\mathcal{D}$};
\node at(0,-0.5) {$\widetilde{\mathcal{O}}(x)$};
\node at(0.5, 0.3){$L$};
\end{scope}

\node at(1.5,0) {$=$};

\end{tikzpicture}
    \caption{Passing the topological defect $\cD$ through $\mathcal{O}$, the point operator will become attached to a line operator $L\subset \cD\otimes\bar{\cD}=A'$. }
    \label{fig: passing D theough O}
\end{figure}

Now, consider a local operator $\mathcal{O}(x)$. When we pass the topological defect $\cD$ through $\mathcal{O}$, seeing Figure \ref{fig: passing D theough O}, the point operator will become attached to a line operator $L$ if it transforms nontrivially under a line operator in $A$. Because $L\subset\cD\otimes\bar{\cD}=A'$, it follows from \eqref{aV=Va} that $L$ must commute with $V_g$. This will be important in examples with non-abelian $G$. 

This already has important consequences for the fermion-boundary scattering process. The scattering process is specified by the defect $\cD$ after folding it back, and $G$ is the symmetry to be preserved by the boundary. This means that the outgoing state can only be twisted by a topological line that \emph{commutes} with the symmetry $G$.

\subsection{Probing $\mathcal{D}$ via Torus Partition Function}

Now we discuss the implication of \eqref{VD=DU} on the torus partition function, which will be useful in discussing various examples.

The existence of a topological interface implies that the theory $\mathcal{T}$ is invariant under gauging $A=\bar{\cD}\otimes\cD$, i.e. 
\begin{eqnarray}\label{eq:TTA}
    \mathcal{T}\simeq \mathcal{T}/A.
\end{eqnarray}
Although the theory is self dual under gauging, various operators may transform non-trivially. In particular, suppose the line operator $U_g$ acts within the theory $\mathcal{T}$, under gauging $A$, it may become a presumably different operator $V_g$ acting within $\mathcal{T}/A\simeq \mathcal{T}$.

\begin{figure}
    \centering
    \begin{tikzpicture}[
    scale=1,
    line join=round,
    line cap=round,
    >=stealth,
    every node/.style={font=\footnotesize}
]

\coordinate (P1) at (0,4.6);
\coordinate (P2) at (4.7,4.6);
\coordinate (P3) at (9.4,4.6);
\coordinate (P4) at (2.1,0);
\coordinate (P5) at (7.2,0);

\begin{scope}[shift={(P1)}]
\draw[thin] (0,0) rectangle (3,3);
\fill[gray!20, even odd rule]
        (0,0) rectangle (3,3)
        (1.5,1.5) circle (0.8);
\draw[
    postaction={decorate},
    decoration={markings, mark=at position 0.125 with {\arrow[scale=0.8]{<}}}
] (1.5,1.5) circle (0.8);
\draw[
    blue,
    decoration={markings, mark=at position 0.4 with {\arrow[scale=0.8]{stealth}}},
    postaction=decorate
] (0.73,1.7) -- (0,1.7);
\draw[
    red,
    decoration={markings, mark=at position 0.4 with {\arrow[scale=0.8]{stealth}}},
    postaction=decorate
] (2.27,1.7) -- (0.73,1.7);
\draw[blue] (2.27,1.7) -- (3,1.7);
\node at (0.38,0.22) {$\mathcal{T}/A$};
\node at (1.5,0.95) {$\mathcal{T}$};
\node[blue] at (0.28,1.95) {$V_g$};
\node[red] at (1.48,1.98) {$U_g$};
\node at (2.03,2.35) {$\mathcal{D}$};
\end{scope}

\begin{scope}[shift={(P2)}]
\draw[thin] (0,0) rectangle (3,3);
\fill[gray!20, even odd rule]
    (0,0) rectangle (3,3)
    (0,3) -- (0,2) arc[start angle=-90,end angle=0,radius=1] -- cycle
    (3,3) -- (2,3) arc[start angle=180,end angle=270,radius=1] -- cycle
    (0,0) -- (1,0) arc[start angle=0,end angle=90,radius=1] -- cycle
    (3,0) -- (3,1) arc[start angle=90,end angle=180,radius=1] -- cycle;
\draw[
    postaction={decorate},
    decoration={markings, mark=at position 0.5 with {\arrow[scale=0.8]{<}}}
] (0,2) arc[start angle=-90,end angle=0,radius=1];
\draw[
    postaction={decorate},
    decoration={markings, mark=at position 0.5 with {\arrow[scale=0.8]{<}}}
] (2,3) arc[start angle=180,end angle=270,radius=1];
\draw[
    postaction={decorate},
    decoration={markings, mark=at position 0.5 with {\arrow[scale=0.8]{<}}}
] (1,0) arc[start angle=0,end angle=90,radius=1];
\draw[
    postaction={decorate},
    decoration={markings, mark=at position 0.5 with {\arrow[scale=0.8]{<}}}
] (3,1) arc[start angle=90,end angle=180,radius=1];
\draw[
    red,
    decoration={markings, mark=at position 0.4 with {\arrow[scale=0.8]{stealth}}},
    postaction=decorate
] (0.87,0.5) -- (0,0.5);
\draw[
    blue,
    decoration={markings, mark=at position 0.4 with {\arrow[scale=0.8]{stealth}}},
    postaction=decorate
] (2.12,0.5) -- (0.87,0.5);
\draw[red] (2.12,0.5) -- (3,0.5);
\node at (0.32,0.22) {$\mathcal{T}$};
\node at (1.5,2.72) {$\mathcal{T}/A$};
\node at (2.18,0.86) {$\mathcal{D}$};
\node[red] at (2.72,0.32) {$U_g$};
\node[blue] at (1.50,0.78) {$V_g$};
\end{scope}

\begin{scope}[shift={(P3)}]
\draw[thin] (0,0) rectangle (3,3);
\fill[gray!20] (1.5,1.5) circle (0.8);

\draw[
    postaction={decorate},
    decoration={markings, mark=at position 0.125 with {\arrow[scale=0.8]{>}}}
] (1.5,1.5) circle (0.8);
\draw[
    decoration={markings, mark=at position 0.4 with {\arrow[scale=0.8]{stealth}}},
    postaction=decorate
] (0,1.5) -- (0.7,1.5);
\draw[
    decoration={markings, mark=at position 0.4 with {\arrow[scale=0.8]{stealth}}},
    postaction=decorate
] (2.3,1.5) -- (3,1.5);
\draw[
    decoration={markings, mark=at position 0.4 with {\arrow[scale=0.8]{stealth}}},
    postaction=decorate
] (1.5,0) -- (1.5,0.7);
\draw[
    decoration={markings, mark=at position 0.4 with {\arrow[scale=0.8]{stealth}}},
    postaction=decorate
] (1.5,2.3) -- (1.5,3);
\draw[
    red,
    decoration={markings, mark=at position 0.4 with {\arrow[scale=0.8]{stealth}}},
    postaction=decorate
] (3,0.5) -- (0,0.5);
\node at (0.30,0.22) {$\mathcal{T}$};
\node at (1.5,1.15) {$\mathcal{T}/A$};
\node[red] at (2.48,0.72) {$U_g$};
\node at (0.28,1.78) {$A$};
\node at (1.72,2.72) {$A$};
\node at (2.18,2.2) {$\mathcal{D}$};
\end{scope}

\begin{scope}[shift={(P4)}]
\draw[thin] (0,0) rectangle (3,3);
\fill[gray!20] (0,0) rectangle (3,3);
\draw[
    blue,
    decoration={markings, mark=at position 0.4 with {\arrow[scale=0.8]{stealth}}},
    postaction=decorate
] (3,1.2) -- (0,1.2);
\node at (0.40,0.22) {$\mathcal{T}/A$};
\node[blue] at (1.50,1.42) {$V_g$};
\end{scope}
\begin{scope}[shift={(P5)}]
\draw[thin] (0,0) rectangle (3,3);
\draw[
    red,
    decoration={markings, mark=at position 0.4 with {\arrow[scale=0.8]{stealth}}},
    postaction=decorate
] (3,0.5) -- (0,0.5);
\draw[
    decoration={markings, mark=at position 0.8 with {\arrow[scale=0.8]{stealth}}},
    postaction=decorate
] (1.5,0) -- (1.5,3);
\draw[
    decoration={markings, mark=at position 0.4 with {\arrow[scale=0.8]{stealth}}},
    postaction=decorate
] (0,1.5) -- (1.5,1.7);
\draw[
    decoration={markings, mark=at position 0.4 with {\arrow[scale=0.8]{stealth}}},
    postaction=decorate
] (3,1.5) -- (1.5,1.3);
\node at (0.30,0.28) {$\mathcal{T}$};
\node[red] at (2.02,0.72) {$U_g$};
\node at (0.28,1.76) {$A$};
\node at (1.72,2.72) {$A$};
\end{scope}

\draw[
    thick,
    decoration={markings, mark=at position 1 with {\arrow[scale=0.8]{stealth}}},
    postaction=decorate
] (2.6,3.7) -- (2.2,4.1);
\draw[
    thick,
    decoration={markings, mark=at position 1 with {\arrow[scale=0.8]{stealth}}},
    postaction=decorate
] (3.55,6.1) -- (4.05,6.1);
\draw[
    thick,
    decoration={markings, mark=at position 1 with {\arrow[scale=0.8]{stealth}}},
    postaction=decorate
] (8.3,6.1) -- (8.8,6.1);
\draw[
    thick,
    decoration={markings, mark=at position 1 with {\arrow[scale=0.8]{stealth}}},
    postaction=decorate
] (9.9,4.1) -- (9.5,3.7);
\node at (6.15,1.5) {$=$};
\node at ($(P4)!0.5!(P5)+(1.5,1.5)$) {$=$};
\end{tikzpicture}
    \caption{A sequence of topological moves, showing the relation on torus partition function with line insertions.}
    \label{fig: topological moves on T2}
\end{figure}
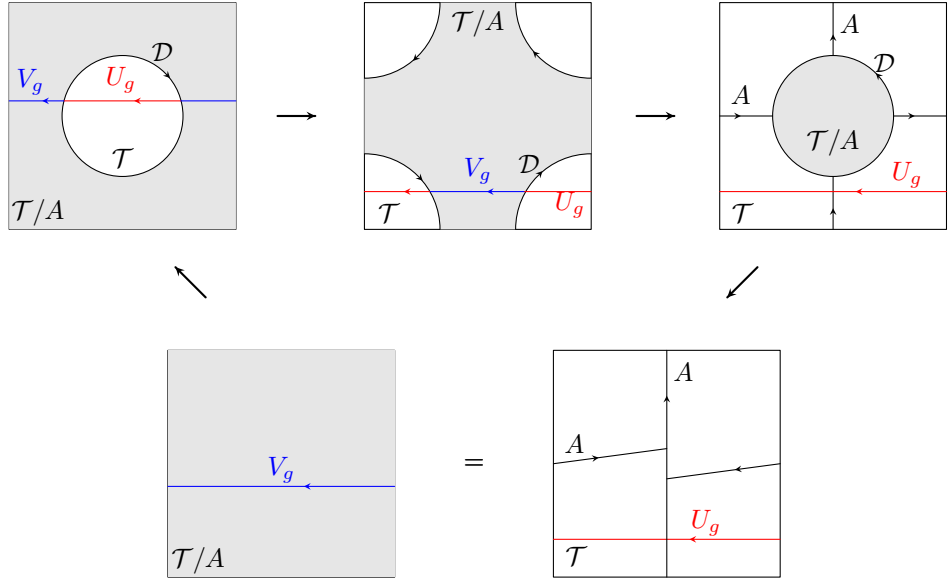

The above discussion implies a relation between the torus partition functions. 
We start by placing $\mathcal{T}/A$ on a torus with an insertion of $V_g$, as shown in Figure \ref{fig: topological moves on T2}, and denote the partition function as $Z_{\mathcal{T}/A}[V_g]$. Then, we blow up a bubble of $\mathcal{T}$ bounded by the interface $\cD$. By shrinking the region on which $\mathcal{T}/A$ is defined and using the relation $\bar{\cD}\otimes\cD=A$, we arrive at the partition function of $\mathcal{T}$ with the insertion of a mesh of $A$ and a line operator $U_g$. Through this sequence of topological moves, we have shown that 
\begin{align}\label{partition func id}
    Z_{\mathcal{T}}[A,U_g]=Z_{\mathcal{T}/A}[V_g] = Z_{\mathcal{T}}[V_g],
\end{align}
where in the second equality, we used the isomorphism \eqref{eq:TTA}.
Note that \eqref{partition func id} is just a schematic expression in which we have omitted the normalization factors coming from the quantum dimension of $\cD$ and the (co)product junctions of $A$. This identity will be a key tool to determine the existence of $\mathcal{D}$.

\section{Abelian Group Intertwining Defect}\label{sec:Defects with Abelian Group Symmetry}

In this section, we provide a general construction for the topological defect $\cD$ with $G=U(1)^n$ being an abelian group in 1+1d Weyl fermions. The existence of $\mathcal{D}$ is ensured by the existence of $U(1)^n$ symmetric conformal boundary conditions of $n$ Dirac fermions \cite{Smith:2019jnh}, upon folding.

\subsection{$U(1)^n$ Symmetric Conformal Boundary Condition}\label{U1nsymmetricBC}
We start by considering $n$ massless Dirac fermions $\Psi^L_i, \Psi^R_i$, $i=1, ..., n$. In \cite{Smith:2019jnh, Smith:2020rru, Smith:2020nuf}, Boyle-Smith and Tong worked out conformal boundary states/conditions preserving the $U(1)^n= U(1)_1 \times ...\times U(1)_n$ chiral symmetry as long as it is anomaly free.  Under the $U(1)_\alpha$, the fermion $\Psi^L_i$ carries charge $Q_{\alpha i}$, and $\Psi^R_i$ carries charge $\tilde{Q}_{\alpha i}$:\footnote{$U_{g_\alpha}$ and $V_{g_\alpha}$ are the same symmetry operator, both associated with $g\in G$, but acting on different Hilbert spaces in different representations of $G$.}
\begin{eqnarray}\label{eq:U1alpha}
U(1)_\alpha: \quad 
    \begin{aligned}
        U_{g_\alpha} \Psi^L_i U_{g_\alpha}^{-1} = e^{-2\pi i Q_{\alpha i} \lambda_\alpha } \Psi^L_i\\
        V_{g_\alpha} \Psi^R_i V_{g_\alpha}^{-1} = e^{-2\pi i \widetilde{Q}_{\alpha i} \lambda_\alpha} \Psi^R_i,
    \end{aligned}
\end{eqnarray}
where $g_\alpha= e^{2\pi i \lambda_\alpha}$. The anomaly of $U(1)^n$ chiral symmetry is given by an integer matrix $k_{\alpha \beta}= k^L_{\alpha\beta} - k^R_{\alpha\beta}$, where $k^L_{\alpha\beta}= \sum_{i}Q_{\alpha i} Q_{\beta i}$ and $k^R_{\alpha\beta}= \sum_{i}\widetilde{Q}_{\alpha i} \widetilde{Q}_{\beta i}$. The $\alpha=\beta$ cases track the self-anomaly of $U(1)_\alpha$, and the $\alpha\neq\beta$ cases track the mixed anomaly of $U(1)_\alpha\times U(1)_\beta$.  The $U(1)^n$ being anomaly free means the charges must satisfy the condition 
\begin{align}\label{QQ=QQ}
    \sum_i Q_{\alpha i}Q_{\beta i}=\sum_i \widetilde{Q}_{\alpha i}\widetilde{Q}_{\beta i}, \quad \forall \alpha, \beta. 
\end{align}

The explicit form of the boundary state $\ket{\mathcal{D}}$ was given by \cite[eq.(2.16)]{Smith:2019jnh}. We will not need its explicit form, apart from the fact that it exists and preserves $U(1)^n$, i.e. 
\begin{eqnarray}\label{eq:UVDD}
    U_{g_\alpha} V^\dagger_{g_\alpha} \ket{\mathcal{D}} = \ket{\mathcal{D}}.
\end{eqnarray}

\subsection{$\mathcal{D}$ is Generically Non-invertible}

Starting from the $U(1)^n$ preserving conformal boundary condition $\ket{\mathcal{D}}$ of $n$ Dirac fermions, we can unfold the theory along the boundary. After unfolding, we get $n$ left moving Weyl fermions $\psi_i$ with a topological defect $\mathcal{D}$. From \eqref{eq:U1alpha}, the $U(1)_\alpha$ acts on the Weyl fermions as
\begin{eqnarray}\label{eq:weylU1alpha}
U(1)_\alpha: \quad 
    \begin{aligned}
        U_{g_\alpha} \psi_i(t,x) U_{g_\alpha}^{-1} = e^{-2\pi i Q_{\alpha i} \lambda_\alpha } \psi_i(t,x), \quad x>0\\
        V_{g_\alpha} \psi_i(t,x) V_{g_\alpha}^{-1} = e^{-2\pi i \widetilde{Q}_{\alpha i} \lambda_\alpha} \psi_i(t,x), \quad x<0.
    \end{aligned}
\end{eqnarray}
Note that the $U_{g_\alpha}$ and $V_{g_\alpha}$ operators does not exist simultaneously in a chiral fermion theory---indeed $U_{g_\alpha}$ only acts on fermions with positive $x$, and $V_{g_\alpha}$ on fermions with negative $x$. This is analogues to the $\bZ_2$ spin flip symmetry and the dual $\bZ_2$ symmetry in the Ising CFT.

For notational simplicity, we use $\psi_i$ to denote the Weyl fermions on the $x>0$ side, and $\widetilde{\psi}_i$ to denote those on the $x<0$ side.

From \eqref{eq:UVDD}, we see that $\mathcal{D}$ must also be $U(1)^n$ symmetric, 
\begin{eqnarray}\label{charges U(1)^n}
    V_{g_\alpha}^\dagger\mathcal{D}U_{g_{\alpha}}= \mathcal{D}, \quad \text{or} \quad V_{g_\alpha} \mathcal{D} = \mathcal{D}U_{g_{\alpha}}.
\end{eqnarray}
In this case, the $U(1)^n$ is anomalous, measured by $k^L_{\alpha\beta}$ and $k^R_{\alpha\beta}$ on the two sides.  The condition \eqref{QQ=QQ} in the Dirac fermion theory amounts to the same anomaly on both sides of $\mathcal{D}$ in the Weyl fermion theory, i.e. $k^L_{\alpha\beta}=k^R_{\alpha\beta}$. Matching of anomaly also ensures that $\mathcal{D}$ is topological, via Sugawara construction.

How to construct $\mathcal{D}$ explicitly? We first show that for generic $Q_{\alpha i}$ and $\widetilde{Q}_{\alpha i}$ satisfying \eqref{QQ=QQ}, $\mathcal{D}$ can not be a unitary operator. We prove by contradiction.

Suppose $\mathcal{D}$ is a unitary operator, \eqref{charges U(1)^n} can be rewritten as 
\begin{eqnarray}
    V_{g_\alpha} = \mathcal{D} U_{g_\alpha} \mathcal{D}^{\dagger}.
\end{eqnarray}
We consider the chiral fermion theory where the $U(1)^n$ charge assignments are as \eqref{eq:weylU1alpha} on the $x<0$ side. Now define $\chi_{i}=\mathcal{D} \widetilde{\psi}_i \mathcal{D}^\dagger$. Since $\mathcal{D}$ is unitary, $\chi_i$ is also a local, chiral, fermion with conformal weight $1/2$. Since all the weight $1/2$ local fermionic operators are spanned by $\widetilde{\psi}_i$ and $\widetilde{\psi}_i^\dagger$, we have
\begin{eqnarray}\label{eq:chiMN}
    \chi_i= \sum_{j=1}^n (M_{ij} \widetilde{\psi}_j+ N_{ij} \widetilde{\psi}^\dagger_j)
\end{eqnarray}
for some matrix $M_{ij}$ and $N_{ij}$. To see its charge under $V_{g_{\alpha}}$, we compute 
\begin{eqnarray}
    V_{g_{\alpha}} \chi_i V_{g_\alpha}^{-1} = \mathcal{D}U_{g_\alpha} \mathcal{D}^\dagger \chi_i \mathcal{D} U_{g_\alpha}^{-1} \mathcal{D}^\dagger= \mathcal{D} U_{g_\alpha} \psi_i U_{g_\alpha}^{-1}\mathcal{D}^\dagger= e^{- 2\pi i Q_{\alpha i} \lambda_{\alpha}} \chi_i
\end{eqnarray}
meaning that $\chi_i$ carries charge $Q_{\alpha i}$. Acting $V_{g_\alpha}$ on both sides of \eqref{eq:chiMN}, one finds $\chi_i = \sum_{j=1}^n e^{2\pi i (Q_{\alpha i}- \widetilde{Q}_{\alpha j}) \lambda_\alpha} M_{ij} \widetilde{\psi}_j + \sum_{j=1}^n e^{2\pi i (Q_{\alpha i}+ \widetilde{Q}_{\alpha j}) \lambda_\alpha} N_{ij} \widetilde{\psi}_j^\dagger$.  Extracting the linear terms of $\lambda_\alpha$ and use the linear independence of $\widetilde{\psi}_j$ and $\widetilde{\psi}_j^\dagger$, we find the following condition 
\begin{eqnarray}
    (Q_{\alpha i}- \widetilde{Q}_{\alpha j})M_{ij}=0, \qquad (Q_{\alpha i}+ \widetilde{Q}_{\alpha j})N_{ij}=0.
\end{eqnarray}
This means 
\begin{eqnarray}
\begin{split}
    &Q_{\alpha i}- \widetilde{Q}_{\alpha j}=0 \quad \text{if} \quad M_{ij}\neq 0\\
    &Q_{\alpha i}+ \widetilde{Q}_{\alpha j}=0 \quad \text{if} \quad N_{ij}\neq 0.
\end{split}
\end{eqnarray}
Among all the matrix elements $M_{ij}, N_{ij}$ for all $i,j=1,...,n$, there is at least one that doesn't vanish. Hence there exists one choice of $i,j$, such that 
\begin{eqnarray}\label{eq:Qproportional}
    Q_{\alpha i}= \widetilde{Q}_{\alpha j} \qquad \text{or} \qquad Q_{\alpha i}=- \widetilde{Q}_{\alpha j}
\end{eqnarray}
for all $\alpha$. This is not satisfied for a generic solution of \eqref{QQ=QQ}. This finishes the proof that $\mathcal{D}$ can not be a unitary, hence invertible, defect in general. 

As a simple example, consider the charge assignment of the 3450 model \cite{Wang:2018ugf,vanBeest:2023dbu,Smith:2019jnh}, where $n=2$ and 
\begin{align}
    Q_{\alpha i}=
    \begin{pmatrix}
        3&4\\
        4&-3
    \end{pmatrix},\quad
    \widetilde{Q}_{\alpha i}=
    \begin{pmatrix}
        5&0\\
        0&5
    \end{pmatrix}.
\end{align}
In this case, the condition \eqref{eq:Qproportional} is clearly not satisfied, so the topological defect $\mathcal{D}$ must be a non-invertible defect.

In fact, the operator $\mathcal{D}$ being a non-invertible defect is anticipated from the folded picture, where a fermion is scattered off by the boundary to a twisted sector operator in many models. In the unfolded setting, $\mathcal{D}$ maps a local operator to a twisted sector operator, which is a feature of non-invertibility.

\subsection{$\mathcal{D}$ is a Duality Defect}
\label{sec:Ddualitydefect}

How to construct the non-invertible defect $\mathcal{D}$ explicitly? Here, we propose that $\mathcal{D}$ is a duality defect associated with gauging certain finite abelian group $\Gamma$. In the special case when $Q$ and $\widetilde{Q}$ come from the unit charge fermion-monopole scattering setting, as shown in \eqref{eq:fermionmonopolecharge}, Ohmori and Watanabe showed that $\mathcal{D}$ is a duality defect associated with gauging $\bZ_n$ or $\bZ_{n/2}$ symmetry, depending on $n$ being odd or even \cite{WatanabeOhmori,Watanabetalk}. Here, we generalize their result to the generic solution of \eqref{QQ=QQ}, by providing a general recipe of determining $\Gamma$ from $Q$ and $\widetilde{Q}$.

Suppose
\begin{eqnarray}\label{ansatz for gamma}
    \Gamma=\bZ_{N_1}\times \cdots \times\bZ_{N_\ell}
\end{eqnarray}
and assume $\mathcal{D}$ is a duality defect obtained by gauging $\Gamma$. The subgroup $\bZ_{N_u}$ acts on the fermion $\psi_i$ with charge $q_{ui}$,
\begin{eqnarray}
    \bZ_{N_u}: \quad \psi_i \to e^{-2\pi i q_{ui}/N_u} \psi_i.
\end{eqnarray}
Both the integers $N_u$ and the charges $q_{ui}$ are to be determined.

As reviewed in Appendix \ref{app: Z of fermion}, the partition function of one Weyl fermion with the twisted boundary conditions $\psi(t,x+2\pi)=-e^{-2\pi ia}\psi(t,x)$ and $\psi(t+2\pi\tau_2,x+2\pi \tau_1)=-e^{-2\pi i b}\psi(t,x)$ is 
\begin{eqnarray}
    Z_{a,b}(\tau) = \frac{1}{\eta(\tau)} \theta
    \begin{bmatrix}
        a\\
        -b
    \end{bmatrix}
    (0,\tau).
\end{eqnarray}
Here $a$ and $b$ can be regarded as the holonomy of a $U(1)$ gauge field along the spatial and temporal cycles, on top of the NS-NS spin structure. The partition function of $n$ Weyl fermions with NS-NS spin structure is simply the product $ Z_{0,0}^n$. 

We assume that the theory of $n$ Weyl fermions is invariant under gauging $\Gamma$. Since gauging $\Gamma$ amounts to summing over all values of holonomies $a_u,b_u\in \bZ_{N_u}$, invariance under gauging means, at the level of torus partition function, 
\begin{eqnarray}\label{Z for gauging}
    \frac{1}{|\Gamma|} \sum_{\{a_u,b_u\}} \prod_{i=1}^n \frac{1}{\eta(\tau)} \theta 
    \begin{bmatrix}
        \sum_{u=1}^{\ell} \frac{q_{ui}}{N_u} a_u\\
        - \sum_{v=1}^{\ell} \frac{q_{vi}}{N_v} b_v
    \end{bmatrix}
    (0,\tau) = 
    \prod_{i=1}^n \frac{1}{\eta(\tau)} \theta
    \begin{bmatrix}
        0\\
        0
    \end{bmatrix}
    (0,\tau),
\end{eqnarray}
where $|\Gamma|=\prod_u N_u$.

Furthermore, the condition \eqref{charges U(1)^n} means that the above equality of partition functions still holds after inserting $U_{g_\alpha}$ and $V_{g_\alpha}$ along the same cycle on the left and right hand side, respectively. 
\begin{eqnarray}\label{eq:keyiden}
    \frac{1}{|\Gamma|}\sum_{\{a_u,b_u\}}\prod_{i=1}^n\frac{1}{\eta(\tau)}\thetab{\sum_{u=1}^\ell \frac{q_{ui}}{N_u}a_u}{-\sum_{v=1}^\ell \frac{q_{vi}}{N_v}b_v-Q_{\alpha i}\lambda_{\alpha}}(0,\tau)=\prod_{i=1}^n\frac{1}{\eta(\tau)}\thetab{0}{-\widetilde{Q}_{\alpha i}\lambda_{\alpha}}(0,\tau).
\end{eqnarray}
We have chosen to insert the $U,V$ defects along the spatial direction. Alternatively, one could choose the other direction subject to adding an appropriate counter term, which we will not discuss for simplicity.

The goal is to find $q_{ui}$ and $N_u$ such that \eqref{eq:keyiden} holds for any $0\leq \lambda_\alpha<1$. This problem can be completely solved. We state the results here and leave the detailed derivation for Appendix \ref{id on thetas}. The gauged group $\Gamma$, defined by $N_u$ and $q_{ui}$, can be determined via the following algorithm. Here we present a more general theorem where we allow  the number of $U(1)$'s, denoted by $N$, to be smaller than the number of fermions $n$, i.e. $N\leq n$. This generalization will be helpful in examples in Section \ref{sec:Nonabelian}. The algorithm is as follows:
\begin{enumerate}
    \item Find a rational orthogonal matrix $R\in O(n,\mathbb{Q})$ such that $\sum_{j=1}^n R_{ij}Q_{\alpha j}=\widetilde{Q}_{\alpha i}$ for all $\alpha=1,...,N$.
    The existence of such an orthogonal matrix $R$ is guaranteed because of \eqref{QQ=QQ}, i.e. $R R^T = R^T  R=1$, but is not unique if $N<n$. When $N=n$, $R$ is unique---the case we mainly focus on.

    \item Choose the smallest integer $d>0$ such that $dR$ is an integer matrix. Compute the Smith normal form of the integer matrix $dR^T$:
    \begin{align}\label{SNF}
        dR^T=USV,
    \end{align}
    where $U,V\in GL(n,\bZ)$ and $S=\operatorname{diag}(s_1,\cdots,s_n)$. Such decomposition is non-unique, and we can use this freedom to bring diagonal elements of $S$ into order $s_1\leq s_2\leq ...\leq s_n$. 

    \item $N_u$ and $q_{ui}$ are then determined by
    \begin{align}\label{N and q}
        N_u=\frac{d}{\gcd(s_u,d)}, \quad q_{ui}=\frac{s_u}{\gcd(s_u,d)} U_{iu}.
    \end{align}
    Any factors yielding $N_u=1$ can be discarded directly, leaving only $\ell$ nontrivial groups, which we relabel as $\prod_{u=1}^\ell\bZ_{N_u}$. 
\end{enumerate}

In summary, the topological defect $\cD$ with the Abelian group symmetry \eqref{charges U(1)^n} exists and is the duality defect obtained by gauging the $\prod_{u=1}^\ell\bZ_{N_u}$ symmetry specified in \eqref{N and q}.\footnote{It can also be $N_u=1$ for all of $u=1,...,n$. In this case, the duality defect is trivial in the sense of gauging an identity group. Nevertheless, it can be a non-trivial invertible defect. See the following paragraph and remarks in Section \ref{sec:remarks}.  }

We emphasize that saying $\mathcal{D}$ is a duality defect does not uniquely specify $\mathcal{D}$. For instance, starting with a vanilla duality defect $\mathcal{D}$, one can compose it with an \emph{abelian} group element $h\in U(1)^n\subset U(n)$ to be $\mathcal{D}':=h\mathcal{D}$. Since $h$ is abelian, $V_g \mathcal{D}'= \mathcal{D}' U_g$ is also satisfied. Indeed, the $U(1)^n$ symmetric conformal boundary condition from \cite{Smith:2019jnh} depends on certain continuous parameter, which should propagate to the definition of $\mathcal{D}$ upon unfolding.

\subsection{Remarks}
\label{sec:remarks}
We make several remarks regarding the above construction.

\subsubsection{Anomaly-Free $\Gamma$}
Since $\Gamma=\prod_{u=1}^\ell \bZ_{N_u}$ is the symmetry we gauge, we must ensure that $\Gamma$ itself is anomaly-free. In fermionic CFT, the chiral anomaly of $\bZ_{N_u}\times \bZ_{N_{v}}$ with charges $q_{ui}$ and $q_{vi}$ is given by\footnote{In bosonic CFT, the self anomaly $k_{uu}$ is modified to be $\sum_{i=1}^n q_{ui}^2 \mod 2N_u$ when $N_u$ is even, and other cases are unmodified. Since we are mainly concerned with fermionic theories, we will not be bothered with this subtlety. }
\begin{align}
    k_{uv}=\sum_{i=1}^n q_{ui}q_{vi} \mod \gcd(N_u,N_v).
\end{align}
In our construction, we assumed an analogue condition, i.e. the lattice condition \eqref{lattice condition}. We note that the lattice condition is stronger than the anomaly-free condition $\sum_{i=1}^n q_{ui}q_{vi}=0$ mod $\gcd(N_u,N_v)$. 

\subsubsection{Absence of $\Gamma-U(1)_\alpha$ Mixed Anomaly}
As illustrated in Figure \ref{fig: topological moves on T2}, there is a 4-way junction between the symmetry line $U_g$ of $U(1)_\alpha$ and the symmetry defect of $\Gamma$. Below, we prove that there is no mixed anomaly between $\Gamma$ and $U(1)_\alpha$. Therefore, such a configuration is unambiguous.

The mixed anomaly between $\bZ_{N_u}$ and $U(1)_\alpha$ is measured by
\begin{align}\label{Z_n U(1) anoamly free}
    \sum_{i=1}^n q_{ui} Q_{\alpha i} \quad \mod N_u.
\end{align}
We rewrite \eqref{SNF} in terms of $q_{ui}$ and $N_u$ defined in \eqref{N and q}\footnote{Here we also include any trivial factors where $N_u=1$.}:
\begin{align}\label{R in q}
    R_{ij}=\sum_{u=1}^n V_{ui}\frac{q_{uj}}{N_u}.
\end{align}
By the definition of $R$,  
\begin{align}
    \widetilde{Q}_{\beta i}=\sum_{j=1}^n R_{ij}Q_{\beta j}=\sum_{u=1}^n\left(\sum_{j=1}^n\frac{q_{uj}Q_{\beta j}}{N_u}\right)V_{ui}\in \bZ.
\end{align}
Since $V\in GL(n,\bZ)$ and $\widetilde{Q}_{\beta i}\in \bZ$, it must be that
\begin{align}
    \sum_{j=1}^n\frac{q_{uj}Q_{\beta j}}{N_u}=(\widetilde{Q}V^{-1})_{\beta u}\in\bZ.
\end{align}
Thus, there is no $\bZ_{N_u}-U(1)_\alpha$ mixed anomaly.

We point out that an alternative way to derive the mixed anomaly condition \eqref{Z_n U(1) anoamly free} in this context is by requiring that the partition function—namely the left-hand side of \eqref{eq:keyiden} with $Q_{\alpha i}\lambda_\alpha$ inserted in the top component of the theta function—is invariant under (large) $\bZ_{N_u}$ gauge transformations:
\begin{align}
    a_u\to a_u+N_u;\quad b_u\to b_u+N_u.
\end{align}
Using the identities listed in Appendix \ref{app: Z of fermion}, one naturally arrives at the requirement $\sum_{i=1}^n q_{ui} Q_{\alpha i}=0 \mod N_u$.

\subsubsection{Quantum Symmetry}
\label{subsubsec:quantumsymmetry}
The theory is invariant under gauging $\prod_{u=1}^\ell \bZ_{N_{u}}$, with charge assigned by $q_{ui}$. After gauging, if the fermion is charged under the $\prod_{u=1}^\ell \bZ_{N_{u}}$, it becomes a twisted sector operator, attached to a quantum symmetry line $L\subset A'=\cD\otimes\bar{\cD}$.

Note that the gauging procedure is completely specified by the matrix $R$. The quantum symmetry can be directly obtained using the exact same algorithm by simply replacing $R$ with $R^{-1}=R^T$.

To discuss more carefully, we work out the symmetry line operator for the quantum symmetry $\widehat{\Gamma}=\prod_{u=1}^\ell \widehat{\bZ}_{N_{u}}$ in the gauged theory. Generally, for group symmetry, the quantum symmetry of the gauged theory $\mathcal{T}/\Gamma$ is specified by the background field $B\in H^1(M,\widehat{\Gamma})$ for $\widehat{\Gamma}=\mathrm{{Hom}(\Gamma,\mathbb{R}/\mathbb{Z})}$. The partition function is
\begin{align}\label{eq:ToverGamma}
    Z_{\mathcal{T}/\Gamma}[B]=\frac{1}{|H^0(M, \Gamma)|}\sum_{a\in H^1(M,\Gamma)}Z_T[A]e^{2\pi i\int a\cup B}.
\end{align}
Thus, the quantum symmetry line is the Wilson line of the gauge field $a$ by taking the background $B$ field as the Poincare dual of some line in spacetime $M$.

We turn on the background $B$ field, taken as a quantum symmetry defect along the time circle, in the left-hand side of \eqref{Z for gauging}, which is
\begin{align}\label{eq:partwithQuantumSym}
    Z_{\mathcal{T}/\Gamma}[B]=\frac{1}{|\Gamma|} \sum_{\{a_u,b_u\}} \prod_{i=1}^n \frac{1}{\eta(\tau)} \theta 
    \begin{bmatrix}
        \sum_{u=1}^{\ell} \frac{q_{ui}}{N_u} a_u\\
        - \sum_{v=1}^{\ell} \frac{q_{vi}}{N_v} b_v
    \end{bmatrix}
    (0,\tau)
    \exp\left(2\pi i \sum_{u=1}^l\frac{b_u}{N_u}B_u\right),
\end{align}
where $B_u\in\bZ_{N_u}=\{0,1,\cdots\}$. Using the identity \eqref{id on theta twisted} proved in Appendix \ref{id on thetas}, we can show that it equals
\begin{align}\label{eq:partwithQuantumSym2}
    Z_{\mathcal{T}/\Gamma}[B]=\prod_{i=1}^n \frac{1}{\eta(\tau)} \theta
    \begin{bmatrix}
        \sum_{u=1}^\ell\frac{V_{ui}B_u}{N_u}\\
        0
    \end{bmatrix}
    (0,\tau).
\end{align}
Thus, the quantum symmetry $\widehat{\bZ}_{N_u}$ acts by
\begin{align}\label{eq:etahataction}
    \widehat{\eta}_u:\quad\widetilde{\psi_i}\to e^{-2\pi i V_{ui}B_u/N_u}\widetilde{\psi_i}.
\end{align}

\subsubsection{When is $\mathcal{D}$ Invertible?}\label{subsec:gaugetrivialsym}
Not every $U(1)^n$ charge assignment $Q_{\alpha i}, \widetilde{Q}_{\alpha i}$ would yield a non-trivial duality defect $\mathcal{D}$ with $V_g \mathcal{D} = \mathcal{D} U_g$. This happens when $N_u=1$ for $u=1,...,n$. 

Note that $N_u=1$ for all $u$ means that $R^T$, hence $R$, is a matrix with integer elements. From the orthogonality of $R$, we have 
\begin{eqnarray}
    \sum_{j=1}^n R_{ij} R_{kj} =\delta_{ik}, \quad i,k=1, ..., n.
\end{eqnarray}
Taking $k=i$, this means that on each row, $(R_{i1}, ..., R_{in})$, there is only one element whose value is $1$ or $-1$, and all others are 0. Taking $k\neq i$, this means that for two different rows, the location of $\pm 1$ are different. Hence $R$ must be a signed permutation matrix. 
Since $\widetilde{Q}_{\alpha i}= \sum_{j=1}^n Q_{\alpha j} R_{ij}$, for any given $i$ there must be certain $j$ with $R_{ij}=\pm 1$, meaning 
\begin{eqnarray}
    \widetilde{Q}_{\alpha i} = \pm Q_{\alpha j}.
\end{eqnarray}
This is precisely the condition of charges, \eqref{eq:Qproportional}, for $\mathcal{D}$ to be invertible.

\subsection{Examples}

We further revisit several examples discussed in the literature \cite{Smith:2019jnh,vanBeest:2023dbu}.

\subsubsection{The 3-4-5-0 Model}
Consider $n=2$ chiral fermions, with $G=U(1)^2$ defined by the following charge assignment:
\begin{align}
    Q_{\alpha i}=
    \begin{pmatrix}
        3&4\\
        4&-3
    \end{pmatrix},\quad
    \widetilde{Q}_{\alpha i}=
    \begin{pmatrix}
        5&0\\
        0&5
    \end{pmatrix}.
\end{align}
Clearly, the condition on $G$ in \eqref{QQ=QQ} is satisfied. Because the number of $U(1)$ factors matches the number of chiral fermions, the matrix $R$ is uniquely fixed:
\begin{align}
    R=\frac{1}{5}
    \begin{pmatrix}
        3&4\\
        4&-3
    \end{pmatrix}.
\end{align}
Applying our algorithm yields $\Gamma=\bZ_5$, with charges $q_i=(3,4)$. The symmetric defect $\cD$ is the duality defect obtained by gauging this $\bZ_5$ group, which has a quantum dimension of $\sqrt{5}$.

Note that the quantum symmetry of this gauging operation is $\bZ_5$ with charges $(3,4)$ acting on the $\widetilde{\psi}_i$'s, which perfectly agrees with the topological line derived in \cite{vanBeest:2023dbu}.

\subsubsection{Scattering off a Unit-Charge Monopole}
Consider $n$ chiral fermions with a symmetry $G=U(1)\times SU(n)$ acting as shown in Table \ref{unite charge monopole}. This is the physical situation deduced from $\text{QED}(\Box,\Box)$ in \cite{vanBeest:2023dbu,Watanabetalk,WatanabeOhmori}.

This scenario is slightly beyond our current scope because $G$ is non-Abelian. However, the Cartan subgroup of $SU(n)$ is $U(1)^{n-1}$, meaning we can obtain a $U(1)^n$ symmetry simply by restricting our attention to the Cartan subgroup of $G$. According to our construction, this restriction is already sufficient to fix the possible defects $\cD$, up to stacking an invertible defect.

\begin{table}[t]
\centering
    \begin{tabular}{c|c|c}
            & $\psi$ & $\widetilde{\psi}$\\
            \hline
            $U(1)$ & 1 & $-1$ \\
            $SU(n)$ & $\Box$ & $\Box$ 
    \end{tabular}
\caption{Symmetries preserved in the unit charge monopole scattering.}
\label{unite charge monopole}
\end{table}

The charges corresponding to $U(1)^n\subset G$ are
\begin{align}\label{eq:fermionmonopolecharge}
Q_{\alpha i}=
\begin{pmatrix}
1 & 1 & 1 & \cdots & 1 \\
1 & -1 & 0 & \cdots & 0 \\
1 & 1 & -2 & \cdots & 0 \\
\vdots & \vdots & \vdots & \ddots & \vdots \\
1 & 1 & 1 & \cdots & -(n-1)
\end{pmatrix},\quad
\widetilde{Q}_{\alpha i}=
\begin{pmatrix}
-1 & -1 & -1 & \cdots & -1 \\
1 & -1 & 0 & \cdots & 0 \\
1 & 1 & -2 & \cdots & 0 \\
\vdots & \vdots & \vdots & \ddots & \vdots \\
1 & 1 & 1 & \cdots & -(n-1)
\end{pmatrix}.
\end{align}
Similarly, the matrix $R$ is completely fixed:
\begin{align}
    R_{ij}=\delta_{ij}-\frac{2}{n}.
\end{align}
Applying the algorithm, we find:
\begin{align}\label{Gamma for n=1 monople}
    \Gamma=
    \left\{
    \begin{array}{cl}
         \bZ_n, & \text{if } n \text{ is odd} \\
         \bZ_{n/2}, & \text{if } n \text{ is even}
    \end{array}
    \right.
    \quad
    q_i=
    \left\{
    \begin{array}{cl}
         (n-2,-2,\cdots,-2), & \text{if } n \text{ is odd} \\
         (\frac{n-2}{2},-1,\cdots,-1), & \text{if } n \text{ is even}.
    \end{array}
    \right.
\end{align}
Note that $\bZ_n$ charges are defined modulo $n$. Here, we leave the explicit $n$ dependence in $q_i$ exactly as it emerges directly from \eqref{N and q}. The gauged group $\Gamma$ acts on all $\psi_i$'s uniformly. Since $R_{ij}=R_{ji}$, the quantum symmetry acts on the $\widetilde{\psi}_i$'s in the exact same manner as shown in \eqref{Gamma for n=1 monople}.

\subsection{Revisiting Fermion-Boundary Scattering}\label{sec: scatter}

Motivated by the existence of $U(1)^n$ symmetric conformal boundary conditions in $n$ Dirac fermions, we have constructed the duality defect $\mathcal{D}$ of $n$ Weyl fermions in this section. Conversely, the duality defect $\mathcal{D}$ gives a more streamlined, symmetry-based, description of the conformal boundary condition upon folding.

The interplay between the conformal boundary condition and topological defect also allows us to compute the fermion-boundary scattering process more systematically. 

Concretely, consider a left Weyl fermion $\psi_i$ moving towards the boundary at $x=0$, we ask what right-moving excitation is  scattered from the boundary. Upon unfolding, this amounts to ask what does $\psi_i$ become when acted upon by $\mathcal{D}$. Since $\psi_i$ carries charge $(q_{1i}, ..., q_{\ell i})$ under $\Gamma=\prod_{u=1}^\ell \bZ_{N_{u}}$ symmetry, after gauging $\Gamma$, the fermion is attached to a topological line 
\begin{eqnarray}
    T_{\{q_{ui}\}}=\prod_{u=1}^n \widehat{\eta}_u^{q_{ui}}
\end{eqnarray}
where $\widehat{\eta}$ is the quantum symmetry line. Here we add up to $n$, including those trivial groups. To determine the $U(1)_\alpha$ symmetry charge in the $T_{\{q_{ui}\}}$-twisted sector vacuum, we refer to the partition function in the $\Gamma$ gauged theory, with $U(1)_\alpha$ defect along the spatial direction, as well as the $T_{\{q_{ui}\}}$ defect along the time direction, turned on. We therefore consider the partition function with background fields $\lambda$ of $U(1)^n$, and $B$ of $\widehat{\Gamma}$, 
\begin{eqnarray}\label{eq:twistp}
    Z_{\mathcal{T}/\Gamma}[B,\lambda] = \prod_{j=1}^n \frac{1}{\eta(\tau)} \theta
    \begin{bmatrix}
        \sum_{u=1}^n \frac{V_{uj}B_u}{N_u}\\
        -\sum_{\alpha=1}^n \widetilde{Q}_{\alpha j} \lambda_\alpha
    \end{bmatrix}(0,\tau).
\end{eqnarray}
Inserting the defect $T_{\{q_{ui}\}}$ amounts to setting $B_u= q_{ui}$.

From the expansion of the theta function as in Appendix \ref{app:twistpartition}, we read off the conformal weight of an arbitrary Virasoro primary state as 
\begin{eqnarray}
    \Delta_{\mathbf{n}}= \frac12 \sum_{j=1}^n \left(n_j + \sum_{u=1}^n \frac{V_{uj}q_{ui}}{N_u}\right)^2.
\end{eqnarray}
Minimizing $\Delta_{\mathbf{n}}$ sets
\begin{eqnarray}
    n_j = \lfloor \frac12 - \sum_{u=1}^n \frac{V_{uj}q_{ui}}{N_u} \rfloor,
\end{eqnarray}
where $\lfloor \cdot \rfloor$ is the floor function. So the state associated with such $n_j$ in the partition function \eqref{eq:twistp} is the vacuum in the $T_{\{q_{ui}\}}$-twist Hilbert space.

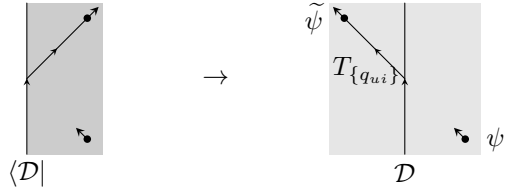
\begin{figure}
    \centering
    \begin{tikzpicture}[
    scale=1,
    line join=round,
    line cap=round,
    every node/.style={font=\footnotesize}
]
\begin{scope}
\fill[gray!40]
        (0,-1) rectangle (1,1);
\draw[decoration = {markings, mark=at position 0.5 with {\arrow[scale=0.8]{stealth}}}, postaction=decorate] (0,-1) -- (0,1);
\node at(0,-1.25){$\bra{\mathcal{D}}$};
\node[circle, fill=black, inner sep=1pt] at (0.8,-0.8) {};
\draw[->, >=stealth, line width=0.4pt] (0.8,-0.8) -- ++(-0.15,0.15);
\node[circle, fill=black, inner sep=1pt] at (0.8,0.8) {};
\draw[->, >=stealth, line width=0.4pt] (0.8,0.8) -- ++(0.15,0.15);
\draw[decoration = {markings, mark=at position 0.5 with {\arrow[scale=0.8]{stealth}}}, postaction=decorate] (0,0) -- (0.8,0.8);
\end{scope}

\node at(2.5,0) {$\to$};

\begin{scope} [xshift=5cm]
\fill[gray!20]
        (-1,-1) rectangle (1,1);
\draw[decoration = {markings, mark=at position 0.5 with {\arrow[scale=0.8]{stealth}}}, postaction=decorate] (0,-1) -- (0,1);
\node at(0,-1.25){$\mathcal{D}$};
\node[circle, fill=black, inner sep=1pt] at (0.8,-0.8) {};
\draw[->, >=stealth, line width=0.4pt] (0.8,-0.8) -- ++(-0.15,0.15);
\node[circle, fill=black, inner sep=1pt] at (-0.8,0.8) {};
\draw[->, >=stealth, line width=0.4pt] (-0.8,0.8) -- ++(-0.15,0.15);
\draw[decoration = {markings, mark=at position 0.5 with {\arrow[scale=0.8]{stealth}}}, postaction=decorate] (0,0) -- (-0.8,0.8);
\node at(1.2,-0.8){$\psi$};
\node at(-1.2,0.8){$\widetilde{\psi}$};
\node at(-0.5,0.1){$T_{\{q_{ui}\}}$};
\end{scope}
\end{tikzpicture}
    \caption{Scattering process}
    \label{fig:scatter}
\end{figure}

The $U(1)_\alpha$ charge of the $T_{\{q_{ui}\}}$-twist vacuum is $Q_{\alpha}^{T\text{vac}}=\sum_{j=1}^n \widetilde{Q}_{\alpha j} (n_j + \sum_{u=1}^n \frac{V_{uj}q_{ui}}{N_u})$. As a consistency check, the charge must be an integer because, using \eqref{R in q}, 
\begin{eqnarray}
    \sum_{j=1}^n \sum_{u=1}^n \widetilde{Q}_{\alpha j} \frac{V_{uj}q_{ui}}{N_u} = \sum_{j=1}^n \widetilde{Q}_{\alpha j} R_{ji} = Q_{\alpha i}\in \bZ.
\end{eqnarray}
So the $U(1)_\alpha$ charge in the twisted sector vacuum is 
\begin{eqnarray}
    Q_{\alpha}^{T\text{vac}}=Q_{\alpha i} + \sum_{j=1}^n \widetilde{Q}_{\alpha j} n_j = Q_{\alpha i} + \sum_{j=1}^n \widetilde{Q}_{\alpha j} \lfloor \frac12 - \sum_{u=1}^n \frac{V_{uj}q_{ui}}{N_u} \rfloor.
\end{eqnarray}
The first piece on the RHS, $Q_{\alpha i}$, precisely matches the charge of the initial state. The second piece, is the offset, which should be compensated by attaching additional local fermions $\prod_{j=1}^n  \widetilde{\psi}_j^{m_j}$\footnote{By $\psi_i^{-1}$ we mean $\psi^\dagger_i$.}. Since these fermions carries charge $\sum_{j=1}^n m_j \widetilde{Q}_{\alpha j}$, to compensate the offset, we require $m_j=  \lceil  \sum_{u=1}^n \frac{V_{uj}q_{ui}}{N_u} -\frac12\rceil$. In other words, we derive the scattering process, as in Figure \ref{fig:scatter}
\begin{eqnarray}
    \psi_i \to T_{\{q_{ui}\}} + \prod_{j=1}^n  \widetilde{\psi}_j^{\lceil  \sum_{u=1}^n \frac{V_{uj}q_{ui}}{N_u} -\frac12\rceil}.
\end{eqnarray}
Here, by $T+\mathcal{O}(x)$ we mean the point operator $\mathcal{O}$ located at one end of the line operator $T$, or by the state-operator map we mean $\mathcal{O}(x)\ket{T}$, where $\ket{T}$ is the $T-$twisted vacuum state \cite{Fukusumi:2021zme,Choi:2024tri,Ebisu:2021acm,Fuchs:1997kt,Fuchs:2000cm,Fuchs:2004dz}. This significantly generalizes the scattering process in 3-4-5-0 model and unit-charge monopole model\cite{vanBeest:2023dbu,WatanabeOhmori,Watanabetalk}.

\section{Remarks on Non-Abelian Group Intertwining Defects}
\label{sec:Nonabelian}

In this section, we apply the consistency conditions \eqref{aU=Ua} and \eqref{aV=Va} to discuss situations where gauging finite abelian groups does or does \emph{not} work.

The method used to solve \eqref{partition func id} does not directly apply to non-Abelian groups $G$ because, due to the torus geometry, the torus partition function only tracks the conjugacy class of $g$. More precisely, we have
\begin{align}
    Z[V_{hgh^{-1}}]=Z[V_hV_gV_{h^{-1}}]=Z[V_g],
\end{align}
and similarly for $Z[A,U_g]$, given that $U_g$ commutes with $A$. This implies that we can only formulate a proposal by restricting to the Cartan subgroup, thereby reducing the problem to the Abelian case discussed previously.

Instead, we turn to the consistency conditions \eqref{aU=Ua} and \eqref{aV=Va} to determine whether gauging finite Abelian groups is a viable approach.

\subsection{Higher-Charge Monopoles}\label{higher charge monopole}
Consider $nN_f$ chiral fermions $\psi_{i\mu}$, where $i=1,\cdots,N_f$ and $\mu=1,\cdots,n$, with the symmetry $G$ specified in Tables \ref{tab:QED box box} and \ref{tab:QED box box bar}. These represent two cases dimensionally reduced from $\text{QED}_4$ by considering fermions scattering off a background monopole of charge $n\geq2$ \cite{vanBeest:2023dbu}. Here, $G$ is the symmetry that could potentially be preserved by the boundary condition induced by the monopole. Upon unfolding, such boundary condition induces the $G$-intertwining topological defect, which we aim to construct. 

\begin{table}[htbp]
    \centering
    \begin{minipage}[t]{0.45\textwidth}
        \centering
        \begin{tabular}{c|c|c}
                & $\psi$ & $\widetilde{\psi}$\\
                \hline
                $U(1)$ & 1 & -1 \\
                $SU(N_f)$ & $\Box$ & $\Box$\\
                $SU(2)$ & $\mathbf{n}$ & $\mathbf{n}$\\
                $\mathbb{Z}_{nN_f}$ & 1 & 1
        \end{tabular}
        \caption{QED$(\Box,\Box)$}
        \label{tab:QED box box}
    \end{minipage}%
    \hfill %
    \begin{minipage}[t]{0.45\textwidth}
        \centering
        \begin{tabular}{c|c|c}
                & $\psi$ & $\widetilde{\psi}$\\
                \hline
                $U(1)$ & 1 & -1 \\
                $SU(N_f)$ & $\Box$ & $\overline{\rule{0pt}{1.5ex}\Box}$\\
                $SU(2)$ & $\mathbf{n}$ & $\mathbf{n}$\\
                $\mathbb{Z}_{n}$ & 1 & 1
        \end{tabular}
        \caption{QED$(\Box,\overline{\rule{0pt}{1.5ex}\Box})$}
        \label{tab:QED box box bar}
    \end{minipage}
\end{table}

We start by focusing on the continuous symmetry $U(1)\times SU(N_f)\times SU(2)$. From \eqref{aU=Ua}, any line operator in the gauged algebra $A$ must commute with $U_{g}$. Note that the operator $U_g$ acts on the fermions $\psi_{i\mu}$ in an irreducible representation of the continuous group in both cases. Because the gauged symmetry $\Gamma$, assumed to be \eqref{ansatz for gamma}, commutes with $SU(2)$ and $SU(N_f)$, by Schur's lemma, it follows that $\Gamma$ can only act on $\psi_{i\mu}$ uniformly. More concretely, the charges of $\Gamma$ must satisfy
\begin{align}\label{equal q1}
    q_{u,i\mu}=q_u \mod N_u, \quad \text{for any }i \text{ and }\mu,
\end{align}
for some integer $q_u$. The same reasoning applies to $V_g$ and the quantum symmetry, implying that the quantum symmetry acts on $\widetilde{\psi}_{i\mu}$ uniformly.

Now we claim that if we assume the $G=U(1)\times SU(N_f)\times SU(2)$-intertwining defect $\cD$ is a duality defect obtained by gauging finite Abelian groups, it must be the one obtained by gauging $\bZ_{nN_f}$ (for odd $nN_f$) or $\bZ_{nN_f/2}$ (for even $nN_f$).

Note that this is not a trivial corollary from the previous section; the Cartan subgroup $U(1)^{N_f+1}\subset G$ does not uniquely fix the matrix $R$, given that $N_f+1<nN_f$. This statement applies to any $G$, as long as the condition \eqref{equal q1} is imposed.

The sketch of the proof is as follows. Since we assume that $A=\bar{\cD}\otimes \cD=\bigoplus_{t\in\Gamma}t$ for a finite Abelian group $\Gamma$, there must exist a matrix $R$ that yields the $N_u$ and $q_{u}$ data via the algorithm presented in the previous section. It turns out that the matrices $R$ satisfying the uniformity condition \eqref{equal q1} can be fully classified, leaving very limited choices:
\begin{align}
    R=
    \left\{
    \begin{array}{ll}
         \text{signed permutation}, & \text{for any } nN_f \\
         \pm(P-\frac{2}{nN_f}J), & \text{if } nN_f\geq 3\\
         \frac{1}{2}H_4 & \text{if }nN_f=4,
    \end{array}
    \right.
\end{align}
where $P$ is a permutation matrix, $J$ is the all-ones matrix, and $H_4$ is the Hadamard matrix. The first option corresponds to gauging the trivial group as shown in Section \ref{subsec:gaugetrivialsym}, while the rest correspond to gauging $\bZ_{nN_f}$ (for odd $nN_f$) or $\bZ_{nN_f/2}$ (for even $nN_f$). The details of this proof are presented in Appendix \ref{special R}.

Now we return to the higher-charge monopole scattering problem.

\paragraph{QED$(\Box,\Box)$}
To satisfy $RQ_\alpha=\widetilde{Q}_\alpha$ for the charges of the Cartan subgroup of $G$, we choose
\begin{align}
    R=I-\frac{2}{nN_f}J.
\end{align}
The defect $\cD$ is the duality defect obtained by gauging $\bZ_{nN_f}$ (for odd $nN_f$) or $\bZ_{nN_f/2}$ (for even $nN_f$). The charges $(q_1, ..., q_{nN_f})$, are $(-2,...,-2)$ mod $nN_f$ for odd $nN_f$, and $(-1,...,-1)$ mod $nN_f/2$ for even $nN_f$.  The quantum symmetry is correspondingly $\widehat{\bZ}_{nN_f}$ or $\widehat{\bZ}_{nN_f/2}$. 

Finally, we check that this proposal also works for the last $\bZ_{nN_f}$ in Table \ref{tab:QED box box}. Let us denote the generator of $\bZ_{nN_f}$ acting on $\psi_{i\mu}$ to be $\eta$, and the generator of $\bZ_{nN_f}$ acting on  $\widetilde{\psi}_{i\mu}$ to be $\widehat{\eta}$. The $\eta$ and $\widehat{\eta}$ should not be confused with generator of $\Gamma$, which depends on the parity of $nN_f$. By including $\bZ_{nN_f}\subset U(1)$, from the general construction of $\mathcal{D}$, we know that 
\begin{align}
    \cD\eta=\widehat{\eta}^{-1}\cD
\end{align}
since the charge of $U(1)$ flips sign. 
However, $\widehat{\eta}^2$ is the quantum symmetry line operator, i.e., $\widehat{\eta}^2\subset A'=\cD\otimes\bar{\cD}$, regardless of whether $nN_f$ is even or odd. Thus,
\begin{align}
    \widehat{\eta}^2\cD=\cD,
\end{align}
meaning that we have 
\begin{align}
    \cD\eta=\widehat{\eta}\cD.
\end{align}

Thus, we find a duality defect $\mathcal{D}$, associated with gauging $\bZ_{nN_f}$ (for odd $nN_f$) or $\bZ_{nN_f/2}$ (for even $nN_f$), that intertwines all the symmetries of the fermions that descend from the QED$(\Box,\Box)$ model. 

\paragraph{QED$(\Box,\overline{\rule{0pt}{1.5ex}\Box})$}
To satisfy $RQ_\alpha=\widetilde{Q}_\alpha$ for the charges of the Cartan subgroup of $U(1)\times SU(N_f)\times SU(2)$, we can only choose
\begin{align}
    R=-I.
\end{align}
We used the fact that $SU(2)$ representations are (pseudo)real, $\mathbf{n}^*\simeq \mathbf{n}$. This means that $\cD$ is simply an invertible defect, which can only be taken to be the charge conjugation of $U(1)$. However, this choice does not preserve the final $\bZ_n$ symmetry in Table \ref{tab:QED box box bar}. To the best of our knowledge, whether a topological defect $\cD$ exists that intertwines the full symmetry in Table \ref{tab:QED box box bar} remains an open question.\footnote{In \cite{vanBeest:2023dbu}, a scattering process $\psi_{i\mu}\to T_i + \widetilde{\psi}_{i\mu}$ was proposed, where the outgoing state is in the $T_i$ twisted sector. This is in contradiction with the property \eqref{aU=Ua} and \eqref{aV=Va}, hence $\mathcal{D}$ does not intertwine $SU(N_f)$, and upon folding, the boundary condition does not preserve $SU(N_f)$.  Relatedly, it is unclear what $T_i$ becomes after an arbitrary (say, infinitesimal) $SU(N_f)$ transformation, since it becomes a sum of topological lines with non-integer coefficients.  }

\subsection{1-5-7-8-9 Model}
Now we consider more exotic non-abelian symmetries, such as the 1-5-7-8-9 model \cite{vanBeest:2023dbu} and some monopoles in the Standard Model with more subtle $SU(2)$ representations \cite{vanBeest:2023mbs}. For convenience of statement, we focus on the former model.

The 1-5-7-8-9 model has $U(1)\times SU(2)$ symmetry. It consists of 15 Weyl fermions $\psi$ transforming under $\boldsymbol{7}_7\oplus \boldsymbol{8}_8$, and 15 Weyl fermions $\widetilde{\psi}$ transforming under $\boldsymbol{1}_1\oplus \boldsymbol{5}_5\oplus \boldsymbol{9}_9$, where bold font $\mathbf{n}$ means $n$-dimensional representation under $SU(2)$, and the subscript $m$ means charge $m$ under the $U(1)$ symmetry. Since the $U(1)\times SU(2)$ anomalies are the same for both $\psi$ and $\widetilde{\psi}$, there is no anomaly obstruction to a $U(1)\times SU(2)$-intertwining topological defect $\mathcal{D}$.\footnote{In the folded setting, anomaly free is not sufficient to guarantee a simple symmetric conformal boundary condition. See \cite{Wei:2025zyd,SMGtalk} for recent discussions. Here, we assume that there exists a $SU(2)\times U(1)$ symmetric boundary condition, and therefore there exists a $SU(2)\times U(1)$-intertwining defect. }

The upshot is that, using \eqref{aU=Ua} and \eqref{aV=Va}, we can show that there is $\emph{no}$ duality defect associated with gauging any finite abelian group, that intertwines the $SU(2)$ representations $\boldsymbol{7}\oplus \boldsymbol{8}$ and $\boldsymbol{1}\oplus \boldsymbol{5}\oplus \boldsymbol{9}$. Note that this no-go result does not require $U(1)$ to be preserved.

The idea is similar to Section \ref{higher charge monopole}. We aim to look for a duality defect of the Weyl fermions associated with gauging a finite abelian group. Such a duality defect is determined by a matrix $R$. We show that the following two requirements on $R$ cannot be satisfied simultaneously:
\begin{enumerate}
    \item the gauged symmetry $\Gamma$ commutes with $SU(2)$ in $\mathbf{1}\oplus\mathbf{5}\oplus\mathbf{9}$ representation, and the quantum symmetry commutes with $SU(2)$ in $\mathbf{7}\oplus\mathbf{8}$ representation;
    \item $R$ satisfy $\sum_{j}R_{ij}Q_j=\widetilde{Q}_i$, where $Q_i$ and $\widetilde{Q}_i$ are charges of the Cartan subgroup of $SU(2)$:
    \begin{align}
    \begin{split}
        &Q_i=
        \left(
        \begin{array}{c|ccccc|ccccccccc}
            0, &2,&1,&0,&-1,&-2,&4,&3,&2,&1,&0,&-1,&-2,&-3,&-4  
        \end{array}
        \right),\\
        &\widetilde{Q}_i=
        \left(
        \begin{array}{ccccccc|cccccccc}
            3, &2,&1,&0,&-1,&-2,&-3,&\frac{7}{2},&\frac{5}{2},&\frac{3}{2},&\frac{1}{2},&-\frac{1}{2},&-\frac{3}{2},&-\frac{5}{2},&-\frac{7}{2} 
        \end{array}
        \right).
    \end{split}
    \end{align}

\end{enumerate}

By Schur's lemma, $\Gamma$ acts on each irreducible block of $SU(2)$ as a constant multiplication, meaning that the charges should satisfy
\begin{align}\label{q confition}
\begin{split}
    &q_{u2}=q_{u3}=\cdots=q_{u6} \mod N_u,\\
    &q_{u7}=q_{u8}=\cdots=q_{u,15} \mod N_u.
\end{split}
\end{align}
Now write $R_{ij}$ in terms of $q_{ui}$ and $N_u$ as \eqref{R in q}, and separate the integer and fractional part of $R$ as
\begin{align}
    R_{ij}=Z_{ij}+C_{ij},
\end{align}
where $Z_{ij}\in\bZ$\footnote{In fact, $Z_{ij}\in\{0,-1\}$ because of the orthogonality of $R$.} and $0\leq C_{ij}<1$. The condition \eqref{q confition} implies that 
\begin{align}
    C_{i1},\quad C_{i2}=C_{i3}=\cdots=C_{i6},\quad C_{i7}=C_{i8}=\cdots=C_{i,15},
\end{align}
for any $i$. However, because the charges of the Cartan subgroup of $SU(2)$ sum to zero, $\sum_{j=1}^{15} C_{ij}Q_j=0$. Thus, we have
\begin{align}
    \sum_{j=1}^{15}R_{ij}Q_j=\sum_{j=1}^{15} Z_{ij}Q_j\in\bZ.
\end{align}
But $\widetilde{Q}_i$ contains half-integers, thus such $R$ is guaranteed not to exist.

Things are similar in the charge 3 monopole discussed in \cite{vanBeest:2023mbs}, where the in-going fermions only carry odd dimensional representations of $SU(2)$, but the out-going fermions carry even dimensional representations.

In summary, neither invertible defects nor duality defects of finite abelian group intertwines the $SU(2)$ symmetry. Searching for a more general non-invertible symmetry that does the job is still an open problem.

\section*{Acknowledgements}

We are grateful to Philip Boyle-Smith, Yichul Choi, Diego Delmastro, Rishi Mouland, Kantaro Ohmori, Brandon Rayhaun, Yuji Tachikawa, and Masataka Watanabe for helpful conversations, and to Andrea Antinucci for helpful feedback on a draft. We also
thank the Kavli Institute for Theoretical Physics (KITP) for hospitality during the program
GenSym25, during which part of this project was done. This research was supported in part
by grant NSF PHY-2309135 to the KITP. The work of Y.Z. is supported by NSFC grant No.12505093 and the starting funds from University of Chinese Academy of Sciences (UCAS) and from the Kavli Institute for Theoretical Sciences (KITS).

\appendix

\section{Theta Function and Partition Function of Weyl Fermion}\label{app: Z of fermion}
Here we record the definition of the theta function and some useful identities, related to large gauge transformations and modular transformations:
\begin{align}
    \thetab{a}{b}(z,\tau)=\sum_{n\in\bZ} \exp\left[\pi i\tau(n+a)^2+2\pi i(z+b)(n+a)\right];
\end{align}
\begin{align}
    \thetab{a}{b}(0,\tau)=\thetab{a+1}{b}(0,\tau)=e^{-2\pi ia}\thetab{a}{b+1}(0,\tau);
\end{align}
and
\begin{align}
    \thetab{a}{b}(0,-\frac{1}{\tau})=(-i\tau)^{\frac{1}{2}}e^{2\pi iab}\thetab{b}{-a}(0,\tau).
\end{align}

The torus partition function of one Weyl fermion, with twist $\psi(t,x+2\pi)=-e^{-2\pi ia}\psi(t,x)$ and $\psi(t+2\pi\tau_2,x+2\pi \tau_1)=-e^{-2\pi i b}\psi(t,x)$, is\footnote{see, for example, chapter 11 of \cite{Hori:2003ic}} 
\begin{align}
    Z_{a,b}(\tau)=\frac{1}{\eta(\tau)}\thetab{a}{-b}(0,\tau).
\end{align}

\section{More on Duality Defects and Quantum Symmetries}

\subsection{Determining $\Gamma$}\label{id on thetas}

In this appendix, we prove the algorithm proposed in Section \ref{sec:Ddualitydefect}. In this section, the theta functions are evaluated at $(0,\tau)$, hence omitted.

Let $R_{ij}$ be a rational orthogonal $n$-by-$n$ matrix. It can be determined from the charge matrices $Q_{\alpha i}$ and $\widetilde{Q}_{\alpha i}$ by solving $\sum_{j=1}^n R_{ij}Q_{\alpha j}=\widetilde{Q}_{\alpha i}$ for all $\alpha=1,...,N$. The solution is unique only when $N=n$. 

The discussion goes in two steps. 
We first show that the identity 
\begin{align}\label{id on theta}
    \frac{1}{\prod_{u=1}^\ell N_u}\sum_{\{a_u,b_v\}}\prod_{i=1}^n
    \thetab{\sum_{u=1}^{\ell} \frac{q_{ui}}{N_u}a_u}{-\sum_{v=1}^{\ell} \frac{q_{vi}}{N_v}b_v-x_i}=\prod_{i=1}^n\thetab{0}{-\sum_{j=1}^n R_{ij}x_j},
\end{align}
holds if and only if the lattice condition
\begin{align}\label{lattice condition}
    \Lambda(\{q_{ui},N_u\})=R^T(\bZ^n)
\end{align}
is satisfied, where the precise definition of the lattice will be given by \eqref{Lambda def}, together with the charge condition \eqref{qq/NN}. Here $a_u,b_u$ are summed over $\{0,1,\cdots,N_u-1\}$, $x_i$ is an arbitrary real number, and $\gcd(q_{u1}, ..., q_{un}, N_u)=1$. We then determine the integers $q_{ui}, N_u$ by solving the lattice condition, and we provide an explicit algorithm to solve them from the matrix $R$.

We prove the first statement by direct calculation. For the reason which will become clear below, we assume the charge condition 
\begin{align}\label{qq/NN}
    \sum_{i=1}^n\frac{q_{ui}q_{vi}}{N_uN_v}\in\bZ.
\end{align}
Note that this condition is stronger than anomaly-free condition for the $\prod_{u=1}^\ell\bZ_{N_u}$ symmetry.   
We start with expanding the theta function on the left-hand side, 
\begin{align}
\begin{split}
    \text{LHS}&=
    \frac{1}{\prod_{u=1}^\ell N_u} \sum_{\{a_u,b_v,n_i\}}\exp\bigg[\sum_{i=1}^n\bigg(\pi i \tau(\sum_{u=1}^\ell\frac{q_{ui}}{N_u}a_u+n_i)^2\\&\hspace{7cm}-2\pi i(\sum_{v=1}^\ell\frac{q_{vi}}{N_v}b_v+x_i)(\sum_{u=1}^\ell\frac{q_{ui}}{N_u}a_u+n_i)\bigg)\bigg]\\
    &=\frac{1}{\prod_{u=1}^\ell N_u} \sum_{\{a_u,n_i\}}\exp\bigg[\sum_{i=1}^n \bigg(\pi i \tau(\sum_{u=1}^\ell\frac{q_{ui}}{N_u}a_u+n_i)^2-2\pi ix_i(\sum_{u=1}^\ell\frac{q_{ui}}{N_u}a_u+n_i)\bigg)\bigg]\\&\hspace{7cm}\times\sum_{\{b_v\}}\exp\left(-2\pi i\sum_{i=1}^n\sum_{v=1}^\ell b_v\frac{q_{vi}}{N_v}n_i\right).
\end{split}
\end{align}
In the second equality, we used the charge condition \eqref{qq/NN} to trivialize a crossing term containing $b_v a_u$.  Further summing over $b_v$ enforces $\sum_{i=1}^n\frac{q_{ui}}{N_u}n_i\in \bZ$. Thus, 
\begin{align}\label{eq:LHS}
\begin{split}
    \text{LHS}&=\sum_{\substack{\{a_u,n_i\}\\ N_u\mid \sum_{i=1}^n q_{ui}n_i}}\exp\bigg[\sum_{i=1}^n \bigg( \pi i \tau(\sum_{u=1}^\ell \frac{q_{ui}}{N_u}a_u+n_i)^2-2\pi ix_i(\sum_{u=1}^\ell\frac{q_{ui}}{N_u}a_u+n_i)\bigg)\bigg]\\
    &=\sum_{\{k_i\}\in\Lambda}\exp\bigg[\sum_{i=1}^n \left(\pi i \tau k_i^2-2\pi ix_ik_i\right)\bigg],
\end{split}
\end{align}
where $\Lambda$ is defined as
\begin{align}\label{Lambda def}
    \Lambda(\{q_{ui},N_u\})
    =\left\{\left(n_i+\sum_{u=1}^\ell \frac{q_{ui}}{N_u}a_u\right)_{i=1,\cdots,n}
    \;\middle|\;
    n_i\in \bZ, a_u\in \bZ_{N_u}, N_u\mid \sum_{i=1}^n q_{ui}n_i\right\}\subset \mathbb{Q}^n.
\end{align}
Note that after simplification, the LHS has the form the theta function, thus matching the form of the RHS. Concretely, the RHS is
\begin{align}\label{eq:RHS}
\begin{split}
    \text{RHS}&=\sum_{\{m_i\}\in\bZ^n}\exp\bigg[\sum_i\left(\pi i\tau m_i^2-2\pi im_i R_{ij}x_j\right)\bigg]\\
    &=\sum_{\{\tilde{m}_i\}\in R^T\bZ^n}\exp\bigg[\sum_i\left(\pi i\tau \tilde{m}_i^2-2\pi i\tilde{m}_i x_i\right)\bigg],
\end{split}
\end{align}
where we redefined $\tilde{m} = R^T m$ in the second line. Since $R$ is an orthogonal matrix, we have $\sum_{i=1}^n m_i^2 = \sum_{i=1}^n \tilde{m}_i^2$. 
Comparing \eqref{eq:LHS} with \eqref{eq:RHS}, one finds LHS and RHS are equal if the lattice condition \eqref{lattice condition} is satisfied.

Now we proceed to determine $\{q_{ui}, N_u\}$ from the lattice condition. The lattice condition implies that the lattice $R^T\bZ^n$ mod 1 are precisely given by integer multiple of $q_{ui}/N_u$. To extract the fractional part of $R^T\bZ^n$, we compute\footnote{One can alternatively understand this formula as follows. We look for group elements $g_{\alpha}=e^{2\pi i \lambda}$, such that $V_{g_\alpha}=1$, i.e. $e^{2\pi i \widetilde{Q}_{\alpha i} \lambda_\alpha}=1$ for any $i$. By the intertwining property \eqref{VD=DU}, $\mathcal{D}=\mathcal{D}U_g$, so these $U_g$ with $g$ determined above form $\Gamma$. More concretely, $\Gamma= \{Q^T\lambda \text{ mod } \bZ^n| \widetilde{Q}^T \lambda \in \bZ^n\}=\{\mathbf{y} \text{ mod } \bZ^n| R^T \mathbf{y}\in \bZ^n\}$, equivalent to \eqref{eq:RmodIntegers}. This idea was used in \cite{Antinucci:2026uuh} to determine $\Gamma=\bZ_5$ in 3450 model. We are grateful to Andrea Antinucci for this observation.  }
\begin{eqnarray}\label{eq:RmodIntegers}
    \Gamma=\frac{R^T\bZ^n}{R^T\bZ^n\cap \bZ^n}.
\end{eqnarray}
Since $R$ is a rational matrix, there exists an integer $d>0$ such that $dR$ is an integer matrix. Let's introduce $W=dR^T$. It is useful to consider the Smith normal form of the integer matrix $W$,
\begin{align}
    W=USV,
\end{align}
where $U,V\in GL(n,\bZ)$ are invertible matrices over the integers, and $S=\operatorname{diag}(s_1,\cdots,s_n)$. Since $V$ is a $GL(n,\bZ)$ matrix, there is $V\bZ^n \simeq \bZ^n$, hence an isomorphism
\begin{align}
\begin{split}
    S\bZ^n&\to W\bZ^n\simeq US\bZ^n\\
    x&\mapsto Ux.
\end{split}
\end{align}
Then \eqref{eq:RmodIntegers} can be simplified as
\begin{align}
    \frac{R^T\bZ^n}{R^T\bZ^n\cap \bZ^n}=\frac{\frac{1}{d}W\bZ^n}{\frac{1}{d}W\bZ^n\cap \bZ^n}\simeq\frac{\frac{1}{d}US\bZ^n}{\frac{1}{d}US\bZ^n\cap \bZ^n}= \frac{U (\oplus_{u=1}^n\frac{s_u}{d}\bZ)}{U (\oplus_{u=1}^n\frac{s_u}{d}\bZ)\cap\bZ^n}.
\end{align}
Thus, we have all the generators
\begin{align}
\begin{split}
    \frac{R^T\bZ^n}{R^T\bZ^n\cap \bZ^n}
    &= \left\langle \frac{s_u}{d}(U_{1u}, ..., U_{nu})^T\bigg|_{u=1,...,n} \; \Bigg|\; \frac{s_u}{d}\notin \bZ\right\rangle.
\end{split}
\end{align}
The lattice condition means that the above generators should be $(\frac{q_{u1}}{N_u}, ..., \frac{q_{un}}{N_u})^T|_{u=1, ..., n}$, with $\gcd(N_u, q_{u1}, ..., q_{un})=1$. Comparing the two expressions of generators, we require $\frac{s_u U_{iu}}{d}= \frac{q_{ui}}{N_u}$. Taking the coprime condition into account, we find\footnote{We should have $N_u=d/L$ and  $q_{ui}=s_u U_{iu}/L$, where $L=\gcd(d,s_uU_{1u}, ..., s_u U_{nu})$. Since $U$ is a unimodular matrix, the element from each column are coprime, i.e. $\gcd(U_{1u}, ..., U_{nu})=1$. Hence $L= \gcd(d, s_u \gcd(U_{1u}, ..., U_{nu}))= \gcd(d,s_u)$, as claimed in \eqref{eq:Nq}. }
\begin{align}\label{eq:Nq}
    N_u=\frac{d}{\gcd(s_u,d)}, \quad q_{ui}=\frac{s_u}{\gcd(s_u,d)} U_{iu}.
\end{align}
When $N_u=1$, we can simply discard the corresponding $u$, and only keep the $u$'s with $N_u>1$, which we arrange as $u=1, ..., \ell$.  

Having determined $\{q_{ui},N_u\}$, let's check that the lattice $\Lambda(\{q_{ui},N_u\})$ with \eqref{eq:Nq} indeed satisfies the lattice condition \eqref{lattice condition}.  Denote $\mathbf{q}_u := (q_{u1}, ..., q_{un})$. Since $\frac{\mathbf{q}_u}{N_u}\in R^T \bZ^n$, it is guaranteed that the assumption \eqref{qq/NN} is satisfied.\footnote{$R^T\bZ^n$, as an orthogonal transformation of $\bZ^n$, has the property that $\mathbf{u}\cdot \mathbf{v}\in\bZ$ for any $\mathbf{u},\mathbf{v}\in R^T\bZ^n$.} Any point in $R^T\bZ^n$ is of the form
\begin{align}\label{qa/N+n}
    \sum_{u=1}^\ell \frac{\mathbf{q}_u}{N_u} a_u+\mathbf{n},\quad \text{where }\mathbf{n}\in\bZ^n, a_u\in\bZ.
\end{align}
The necessary and sufficient condition for $\sum_u \frac{\mathbf{q}_u}{N_u} a_u+\mathbf{n}\in R^T\bZ^n$ is that 
\begin{align}
    \left(\sum_{u=1}^\ell \frac{\mathbf{q}_u}{N_u} a_u+\mathbf{n}\right)\cdot \mathbf{v}\in \bZ
\end{align}
for any $\mathbf{v}\in R^T\bZ^n$. Since $\frac{\mathbf{q}_u}{N_u}\in R^T\bZ^n$, $\frac{\mathbf{q}_u}{N_u}\cdot \mathbf{v}\in\bZ$, we only require that $\mathbf{n}\cdot \mathbf{v}\in\bZ$. But $\mathbf{v}$ is also of the form \eqref{qa/N+n}, which we denote as $\mathbf{v}= \sum_{u=1}^\ell \frac{\mathbf{q}_u}{N_u} b_u + \mathbf{m}$. Then $\mathbf{n}\cdot \mathbf{v}\in\bZ$ is equivalent to the condition
\begin{align}
    \frac{\mathbf{q}_u}{N_u}\cdot \mathbf{n}=\sum_{i=1}^n \frac{q_{ui}}{N_u}n_i\in\bZ,
\end{align}
which recovers the condition in the definition of $\Lambda(\{q_{ui}, N_u\})$ \eqref{Lambda def}. Thus, we has proved the lattice condition \eqref{lattice condition}.

\subsection{Adding Twists of Quantum Symmetry}
\label{app:twistpartition}

In the discussion of quantum symmetry in Section \ref{sec:remarks} and \ref{sec: scatter}, we utilize a twisted version of the identity \eqref{id on theta}:
\begin{align}\label{id on theta twisted}
    \frac{1}{\prod_{u=1}^\ell N_u}\sum_{\{a_u,b_v\}}\prod_{i=1}^n
    \thetab{\sum_{u=1}^{\ell} \frac{q_{ui}}{N_u}a_u}{-\sum_{v=1}^{\ell} \frac{q_{vi}}{N_v}b_v-x_i}\exp\left(2\pi i \sum_{u=1}^l\frac{b_u}{N_u}B_u\right)
    =\prod_{i=1}^n\thetab{\sum_{u=1}^\ell\frac{V_{ui}B_u}{N_u}}{-\sum_{j=1}^n R_{ij}x_j}.
\end{align}

The proof is almost parallel, but this time summing over $b_v$ enforces $\sum_{i=1}^n\frac{q_{ui}}{N_u}n_i-\frac{B_u}{N_u}\in \bZ$. Thus, 
\begin{align}\label{eq:LHS twisted}
    \text{LHS}=\sum_{\{k_i\}\in\Lambda+\{n^*_i\}}\exp\bigg[\sum_{i=1}^n \left(\pi i \tau k_i^2-2\pi ix_ik_i\right)\bigg],
\end{align}
where $n^*_i\in\bZ$ is a solution to
\begin{align}
    \sum_{i=1}^nq_{ui}n^*_i=B_u\mod N_u.
\end{align}
Because of the \eqref{lattice condition},
\begin{align}\label{eq:Rijnj}
\begin{split}
    \text{LHS}&=\sum_{\{m_i\}\in\bZ^n}\exp\bigg[\pi i \tau (\mathbf{m}+R\mathbf{n^*})^2-2\pi i(R\mathbf{x})^T\cdot(\mathbf{m}+R\mathbf{n^*})\bigg]=\prod_{i=1}^n\thetab{\sum_{j=1}^nR_{ij}n^*_j}{-\sum_{j=1}^n R_{ij}x_j}.
\end{split}
\end{align}
From the definition of $q_{ui}$ \eqref{N and q}, we have
\begin{align}
    B_u=\frac{s_u}{\gcd(s_u,d)}\sum_{i=1}^nU_{iu}n_i^*\mod N_u.
\end{align}
Then it follows that
\begin{align}
    \sum_{j=1}^nR_{ij}n^*_j=\sum_{j,u=1}^n \frac{1}{d} V_{ui} s_u U_{ju} n_j^* \stackrel{\text{mod } 1}{=} \sum_{u=1}^n \frac{1}{d}V_{ui} \gcd(s_u,d) B_u =\sum_{u=1}^\ell\frac{V_{ui}B_u}{N_u}\mod 1.
\end{align}
We suppressed the terms with $N_u=1$. Substituting it into \eqref{eq:Rijnj}, we get the claimed identity \eqref{id on theta twisted}.

\section{Classification of a Special Kind of Matrix $R$}\label{special R}
In this section, we prove the following statement. Let $R$ be an $N$-by-$N$ rational orthogonal matrix, $q_{ui}$ and $N_u$ be the integers obtained from $R$ by the algorithm in Appendix  \ref{id on theta}, and $q'_{ui}$ and $N'_u$ be the ones obtained from $R^T$. By requiring that
\begin{align}\label{equal q}
\begin{split}
    &q_{ui}=q_u\mod N_u\\
    &q'_{ui}=q'_u\mod N'_u,
\end{split}
\end{align}
all possible matrices $R$ are classified by
\begin{align}\label{summary of R}
    R=
    \left\{
    \begin{array}{ll}
         \text{signed permutation}, & \text{for any } N \\
         \pm(P-\frac{2}{N}J), & \text{if } N\geq 3\\
         \frac{1}{2}H_4 & \text{if }N=4,
    \end{array}
    \right.
\end{align}
where $P$ is a permutation matrix, $J$ is the all-ones matrix, and $H_4$ is the Hadamard matrix.

We rewrite $R_{ij}$ in terms of $q_{ui}$ and $N_u$ as in \eqref{R in q}. The first equation in \eqref{equal q} then implies that 
\begin{align}
    R_{ij_1}=R_{ij_2} \mod 1.
\end{align}
By separating $R$ into an integer part and a fractional part,
\begin{align}
    R_{ij}=Z_{ij}+C_{ij},
\end{align}
where $Z_{ij}\in\bZ$ and $0\leq C_{ij}<1$, we obtain 
\begin{align}
    C_{i1}=C_{i2}=\cdots=C_{iN}.
\end{align}

Similarly, we rewrite $R_{ij}$ in terms of $q'_{ui}$ and $N'_u$:
\begin{align}
    R_{ij}=\sum_{u=1}^n U_{ju}\frac{q'_{ui}}{N'_u}.
\end{align}
The second equation in \eqref{equal q} implies that 
\begin{align}
    R_{i_1j}=R_{i_2j}\mod 1,
\end{align}
and hence
\begin{align}
    C_{1j}=C_{2j}=\cdots=C_{Nj}.
\end{align}
Together, these two conditions mean that $C_{ij}$ must be a constant matrix, $C_{ij}=C$. The $R$ matrix is simplified to $R=Z+CJ$. 

If $C=0$, then $R=Z$ is a signed permutation matrix, which corresponds to the trivial gauging. See Section \ref{subsec:gaugetrivialsym}. In the following, we consider $0<C<1$. Since $R$ is orthogonal, $|R_{ij}|\leq1$, meaning $Z_{ij}$ can only take values in 
\begin{align}
    Z_{ij}\in \{0,-1\}.
\end{align}
From the orthogonality condition $(Z+CJ)(Z^T+CJ)=I$, we have 
\begin{align}\label{eq on Z}
    C^2N+\sum_{k=1}^N \left(Z_{ik}Z_{jk}+CZ_{ik}+CZ_{jk}\right)=\delta_{ij}.
\end{align}
Considering the diagonal elements ($i=j$) and letting $k_i$ denote the number of $(-1)$ entries in row $i$ of $Z$, we obtain 
\begin{align}\label{eq:kc}
    k_i(1-2C)=1-C^2N.
\end{align}
If $C=\frac{1}{2}$, we find $N=4$ and up to conjugation by signed permutation matrix, $R=\frac{1}{2}H_4$, where $H_4$ is the Hadamard matrix. If we assume $C\neq \frac{1}{2}$, then $k_i=(1-C^2N)/(1-2C)$ is independent of $i$, which we will simply denote by $k$.

Similarly, from $(Z^T+CJ)(Z+CJ)=I$, we can show that the number of $(-1)$ entries in column $i$ of $Z$ is also $k$. Then, summing over $i$ on both sides of \eqref{eq on Z}, gives  
\begin{eqnarray}
    (CN-k)^2=1.
\end{eqnarray}
Combining with \eqref{eq:kc}, we arrive at
\begin{align}
    (k-1)(k-N+1)=0.
\end{align}

If $k=1$, then $C=\frac{2}{N}$. We require $N\geq 3$ in order for $0<C<1$. From \eqref{eq on Z} we have $Z^TZ=I$. Combining this with the fact that $Z$ contains exactly one $-1$ in every row and column, we conclude that $Z=-P$ for some permutation matrix $P$. Thus, $R=-P+\frac{2}{N}J$.

If $k=N-1$, then $C=\frac{N-2}{N}$, which also requires $N\geq 3$. Similarly, from \eqref{eq on Z} we have $(Z+J)^T(Z+J)=I$. Since $Z+J$ contains $1$ exactly once in every row and column, we have $Z+J=P$. Thus, $R=P-\frac{2}{N}J$. These results are summarized in \eqref{summary of R}.

\bibliographystyle{JHEP}
\bibliography{ref}

\end{document}